\documentclass[manuscript, screen, review=false]{acmart}
\renewcommand\footnotetextcopyrightpermission[1]{} 
\AtBeginDocument{%
  }

\usepackage{hyperref}
\usepackage[american]{babel}
\usepackage[autostyle=true,threshold=0,autopunct=true]{csquotes}
\usepackage{booktabs}

\usepackage{tcolorbox}
\usepackage{tikz}
\usepackage{wasysym}
\usepackage{subcaption}
\usepackage{svg}
\usepackage{tabularx}
\usepackage{array}
\usepackage{soul}

\usepackage{geometry} 

\newcommand\TODO[1]{\textcolor{red}{#1}}

\begin{document}

\title{Policy-as-Prompt: Rethinking Content Moderation in the Age of Large Language Models}

\author{Konstantina Palla}
\email{konstantinap@spotify.com}
\affiliation{%
  \institution{Spotify}
  \city{London}
  \country{UK}
}

\author{Jos\'e Luis Redondo Garc\'ia}
\email{joseluisr@spotify.com}
\affiliation{%
  \institution{Spotify}
  \country{Spain}
}

\author{Claudia Hauff}
\email{claudiah@spotify.com}
\affiliation{%
  \institution{Spotify}
  \country{Netherlands}
  }

\author{Francesco Fabbri}
\email{francescof@spotify.com}
\affiliation{%
  \institution{Spotify}
  \country{Spain}
}

\author{Henrik Lindstr\"om}
\email{henok@spotify.com}
\affiliation{%
  \institution{Spotify}
  \country{Sweden}
}

\author{Daniel R. Taber}
\email{dtaber@spotify.com}
\affiliation{%
  \institution{Spotify}
  \country{US}
}

\author{Andreas Damianou}
\email{andreasd@spotify.com}
\affiliation{%
  \institution{Spotify}
  \country{UK}
}

\author{Mounia Lalmas}
\email{mounial@spotify.com}
\affiliation{%
  \institution{Spotify}
  \country{UK}
}


\renewcommand{\shortauthors}{Palla et al.}

\begin{abstract}
 Content moderation plays a critical role in shaping safe and inclusive online environments, balancing platform standards, user expectations, and regulatory frameworks. Traditionally, this process involves operationalising policies into guidelines, which are then used by downstream human moderators for enforcement, or to further annotate datasets for training machine learning moderation models. However, recent advancements in large language models (LLMs) are transforming this landscape. These models can now interpret policies directly as textual inputs, eliminating the need for extensive data curation. This approach offers unprecedented flexibility, as moderation can be dynamically adjusted through natural language interactions. This paradigm shift raises important questions about how policies are operationalized and the implications for content moderation practices.
In this paper, we formalise the emerging policy-as-prompt framework and identify five key challenges across four domains: \textbf{Technical Implementation} (1. translating policy to prompts, 2. sensitivity to prompt structure and formatting), \textbf{Sociotechnical} (3. the risk of technological determinism in policy formation), \textbf{Organisational} (4. evolving roles between policy and machine learning teams), and \textbf{Governance} (5. model governance and  accountability). Through analysing these challenges across technical, sociotechnical, organisational, and governance dimensions, we discuss potential mitigation approaches. This research provides actionable insights for practitioners and lays the groundwork for future exploration of scalable and adaptive content moderation systems in digital ecosystems.
\end{abstract}


\maketitle

\section{Introduction}
Content moderation involves the systematic monitoring and regulation of user-generated content in online platforms. This critical practice fosters safe, inclusive, and respectful digital spaces while aligning with the values of hosting organisations and adhering to regulatory requirements. 
As digital ecosystems grow and content creation surges, the demand for scalable and adaptive moderation systems has become increasingly urgent. Online platforms now face the dual challenge of managing vast, dynamic content streams and meeting evolving societal and regulatory expectations. 
With the internet continuing to transform how we access information, entertainment, and services, content moderation has become essential for effectively managing online platforms, bridging local and global perspectives to address societal needs and foster community trust. 

The architecture of moderation pipelines varies significantly across digital services, reflecting their unique operational needs. For instance, video streaming platforms often prioritise automated detection of copyrighted or harmful content, while community forums like Reddit~\cite{reddit_content_policy} emphasise moderating user interactions and discussions, often requiring human oversight alongside automated tools~\cite{Gorwa2020}. Despite these platform-specific differences, successful content moderation fundamentally relies on two interconnected processes: \emph{operationalisation} and policy \emph{enforcement} \cite{Singhal_2023}. 

\textit{Operationalisation} involves transforming abstract policy objectives into actionable protocols for consistent implementation by human moderators and algorithms. This process includes several critical components: developing comprehensive annotation guidelines, curating high-quality training datasets, setting precise model predictions thresholds, and conducting thorough annotator training. Through operationalisation, abstract policies evolve into specific rules supported by robust workflows and systems. 
Content moderation policies are implemented through various documented forms, which we refer to as policy artifacts. These artifacts operate at different levels of granularity within an organization to cover its moderation needs. At the highest level, platform-wide rules establish foundational principles that guide moderation strategies. These principles then inform more granular artifacts, such as product-specific policies and detailed annotation guidelines, which address domain-specific requirements. This hierarchical structure ensures policies remain broadly applicable while allowing contextual adaptability. 

\textit{Enforcement} puts these artifacts into practice, guiding decision-making processes for both human reviewers and algorithmic systems. During enforcement, content is systematically evaluated against context-appropriate policy artifacts, translating the groundwork laid during operationalisation into concrete moderation actions. Operationalisation provides the foundation for consistent, clear enforcement and works in tandem with the enforcement process. 
The interplay between operationalisation and enforcement is dynamic and interdependent, with changes or inconsistencies in one often impacting the other. Policy artifacts are regularly updated to address emerging content scenarios, refine enforcement nuances, or establish additional boundaries. For example, the appearance of new content types or edge cases may require adjustments to annotation guidelines, ensuring both moderators and algorithms remain aligned with overarching principles.

Technology plays a crucial role in advancing both operationalisation and enforcement, enabling these processes to adapt and remain effective in response to evolving challenges. As the exponential growth of content volumes has driven unprecedented demands on moderation systems, significant technological advancements have emerged to meet these needs. Platforms increasingly rely on automated solutions, with machine learning algorithms becoming central to the moderation pipeline~\cite{Prem2024}. 
This work focuses on a transformative development in AI-assisted enforcement:
the ability to encode content moderation policies directly as natural language prompts in LLMs, a paradigm we term \textbf{``policy-as-prompt''} and show in Figure \ref{fig:policy_prompt_setup}. By eliminating the need for extensive manual annotation pipelines, this approach offers unparalleled flexibility and scalability for moderation systems,  enabling rapid policy iteration and fine-grained control over enforcement decisions. This emerging approach is gaining significant traction - enforcement strategies continue to vary across platforms, but the adoption of AI-driven methods, particularly those leveraging LLMs with injectable guidelines, is rapidly redefining the content moderation landscape~\citep{inan2023llama, Markov2023}. 

To the best of our knowledge, this work presents the first comprehensive exploration of the policy-as-prompt paradigm, examining its key challenges and implications across technical, sociotechnical, organisational, and regulatory domains. Beyond identifying these challenges, we systematically analyse their interdependencies and implications, providing a foundation for further inquiry into this transformative approach.
In Section \ref{sec:the_shift} we examine the technological evolution of content moderation, tracing its progression from basic pattern matching to modern LLM-driven approaches. We explore how this progression has fundamentally altered moderation architectures and introduced the emerging policy-as-prompt. In Sections \ref{sec:technical_implementation_challenges} to \ref{sec:governance_challenges} we break down five key challenges across four domains as outlined in Table~\ref{tab:area_challenges}; \textbf{Technical Implementation} (focusing on the challenges of 1. converting policies to prompts and 2. addressing prompt structure and format sensitivity, which we support with empirical findings), \textbf{Sociotechnical} (3. highlighting the impact of technological determinism in policy formation), \textbf{Organisational} (4. examining the evolving roles of policy-ML teams), and \textbf{Governance} (considering the 5. model governance and accountability). We provide a list of recommendations towards mitigating the challenges in Section \ref{sec:mitigations}. 
We discuss the limitations of this work along with  concluding remarks in Section \ref{sec:conclusion}.



\begin{table}[htb!]
    \centering
    \caption{Challenges across different areas in the `policy-as-prompt' implementation. }
    \label{tab:area_challenges}
    \begin{tabular}{l}
        \toprule
        \textbf{Area \& Challenges} \\
        \midrule
        \textbf{Technical Implementation}   \\
        {1} Converting policies to prompts \\
        {2} Prompt structure and format sensitivity \\
        \midrule
        \textbf{Sociotechnical}    \\
        {3} Technological determinism in policy formation  \\
        \midrule
        \textbf{Organizational}   \\
        {4} Evolving policy-ML team roles \\
        \midrule
        \textbf{Governance} \\
        {5} Model governance and accountability    \\
        \bottomrule
\end{tabular}
\end{table}


\section{The Evolution of Content Moderation Technology}
\label{sec:the_shift}
The evolution of content moderation, particularly within the text domain, has been closely tied to the exponential growth of user-generated content, driven by the rapid advancement of algorithmic capabilities. Early moderation efforts relied on basic automation, functioning like a digital ``find and replace’’ tool. Methods such as keyword filters and hash-matching were used to identify and remove prohibited content~\cite{engstrom2017limits}. While these first-generation tools were effective at detecting exact matches,  they were easily circumvented by minor alterations, underscoring their limitations in adapting to the dynamic nature of online expression~\cite{Reda2017}.

The rapid increase in user-generated content demanded more robust and scalable solutions, driving a shift towards machine learning-based approaches~\cite{Prem2024}. Early word embedding models like Word2Vec~\cite{Mikolov2013} and GloVe~\cite{pennington2014glove} enabled moderation approaches to capture semantic relationships between words, enhancing their ability to detect variations in harmful phrases and toxic language. This marked a pivotal step in moving beyond simple pattern matching to understanding the meaning behind online content. 

However, the true breakthrough came with the introduction of contextual embedding models~\cite{peters-etal-2018-deep} and the transformative Transformer architecture~\cite{Vaswani2017}. Models like BERT~\cite{devlin-etal-2019-bert} enhanced content moderation  by analysing content in context, enabling systems to improve detection of nuances such as sarcasm, coded language, and implicit bias that were previously undetectable. This advancement significantly improved the ability to  understand the intent and impact of online communication, specifically in the realm of textual content~\cite{belloni2023multilingual}. 

Today, advanced language models like GPT series~\cite{openai2023gpt4}, LLaMA series~\cite{touvron2023llama2} and Claude series~\cite{anthropic2023claude} are redefining the boundaries of content moderation. These models depart from traditional supervised learning--where systems learn from training examples-- and are able to interpret desired behaviour directly from textual instructions. Their natural language understanding and reasoning capabilities allow them to go beyond simply identifying harmful content—they can assess its context and potential impact. They showcase capabilities such as applying complex policy guidelines, engaging in nuanced dialogue with users, and adapting to evolving language and online trends~\cite{openai2024moderation, anthropic_content_moderation, desai2024genaiusersafetysurvey, kolla2024, kumar2024, mullick-etal-2023-content, aldahoul2024}. 
%
Rather than relying solely on rigid classification models, content moderation now incorporates general-purpose reasoning systems that provide more flexible and context-sensitive oversight. This approach allows for greater adaptability in addressing emerging challenges, making moderation more effective and responsive to the complexities of online discourse.

By enabling the direct integration of  policy guidelines into prompts, LLMs are reshaping the structure of moderation workflows. In the next section, we explore  the details of this policy-as-prompt setup, examining its impact on workflow design and comparing it with traditional frameworks to provide a clearer context for this paradigm shift. 

\subsection{The Traditional Algorithmic Moderation Pipeline}

\begin{figure}[h]
  \centering
  \begin{subfigure}{0.45\linewidth}
    \centering
    \includegraphics[width=0.8\linewidth]{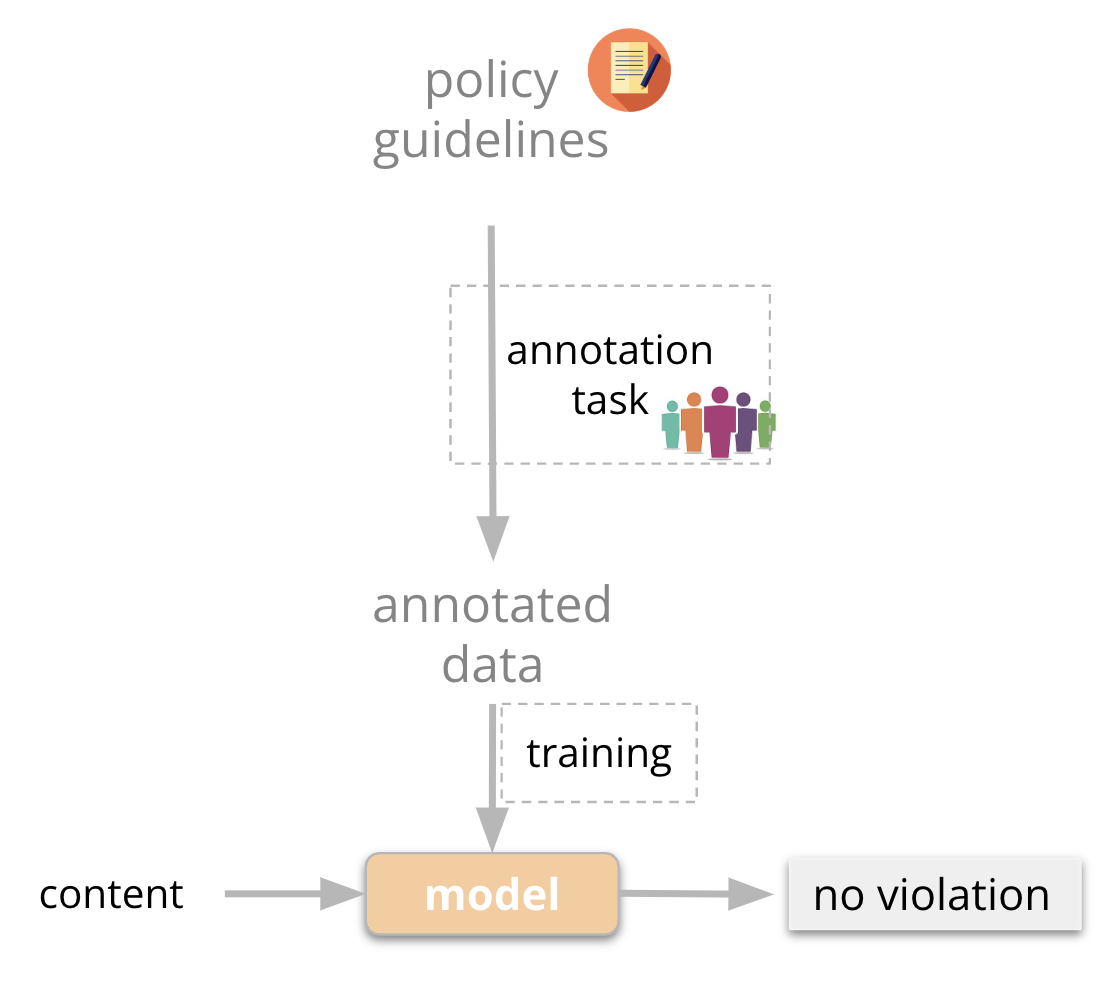}
    \caption{}
    \label{fig:trad_setup}
  \end{subfigure}
  \hspace{0.01in}
  \begin{subfigure}{0.5\linewidth}
    \centering
      \includegraphics[width=\linewidth]{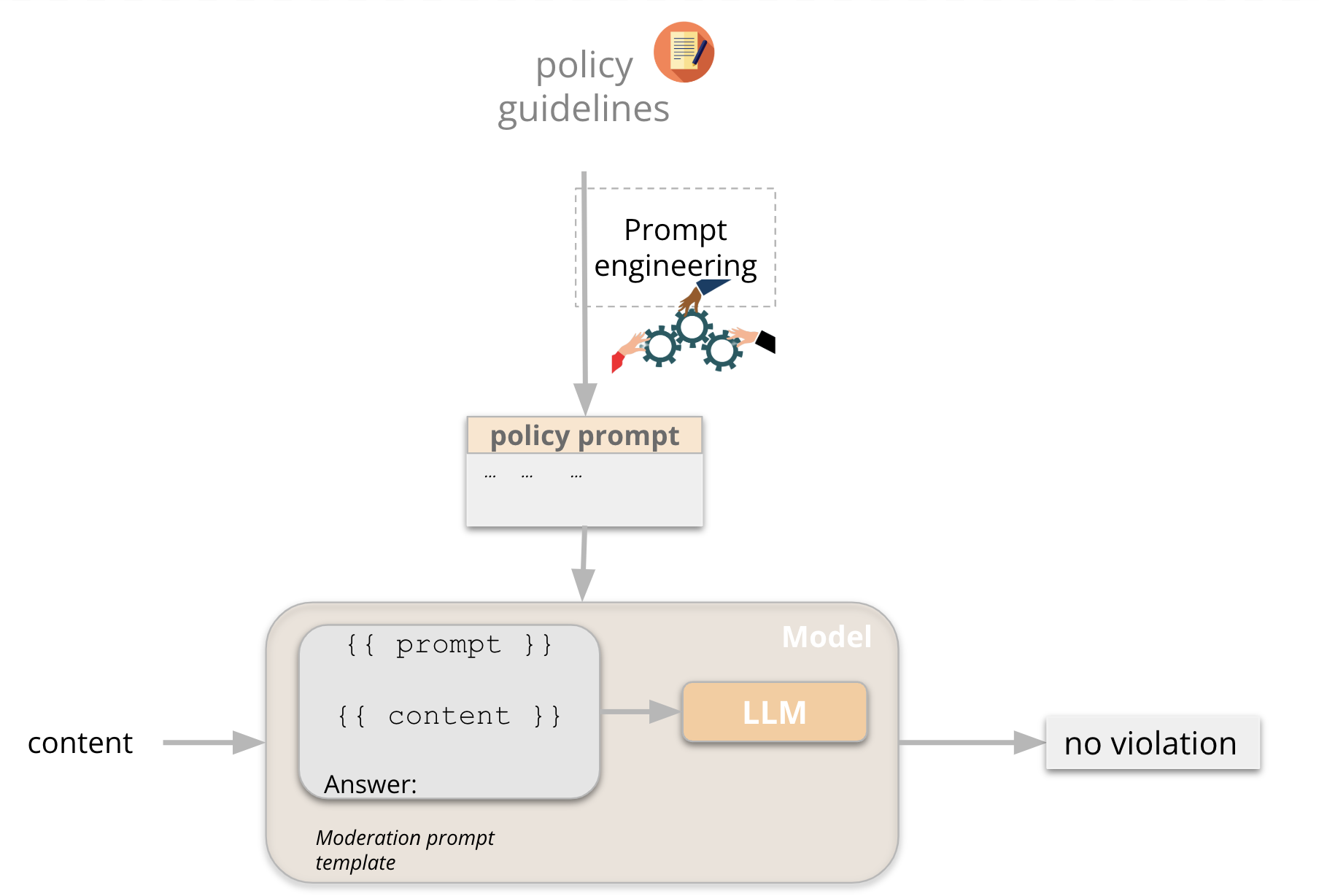} 
    \caption{}
    \label{fig:policy_prompt_setup}
  \end{subfigure}
  \caption{Approaches to content moderation. (a) \textbf{Traditional pipeline}: policy guidelines inform human annotation, which produces training data for models. (b) \textbf{Policy-as-prompt}: policy guidelines are encoded directly as prompts, enabling LLMs to perform moderation without explicit annotation datasets. 
  }
  \label{fig:comparison_diagrams}
  \Description{}
\end{figure}
The adoption of the ML-based approaches required the development of extensive infrastructures for data collection, labelling, model training, and deployment. These requirements fundamentally transformed how platforms approached content moderation at an architectural level.  A standard framework has emerged that integrates human expertise with machine learning models, creating a scalable and widely adopted industry  (see Figure~\ref{fig:comparison_diagrams}a). At the core of this framework are human annotators and machine learning models trained on annotated datasets~\cite{Gillespie2018, Gorwa2020}. Human experts label content based on predefined guidelines, producing datasets that encapsulate these guidelines. 
These annotated datasets are then used to train machine learning models, enabling them to assess whether content violates platform policies. During training, the model adjusts its internal parameters--known as weights--to better align with the patterns identified in the annotated data. Once trained, the model can be deployed to evaluate new, real-world content, autonomously flagging potential policy violations. 

This process is inherently iterative, with models continuously refined based on feedback and new data. However, real-world applications bring additional complexities. Annotators may disagree on how specific content should be labelled, leading to inconsistencies or ambiguities within the dataset~\cite{sandri-etal-2023-dont, larimore-etal-2021-reconsidering}. Furthermore, guideline may undergo revisions, but retrospective relabelling of existing data is often impractical, complicating the training process. Despite these challenges, this integration of human expertise and machine learning has become a foundational approach to scalable content moderation.

\subsection{Policy-as-prompt}
While the annotated dataset approach has long been the industry standard in moderation pipelines with an algorithmic component, the emergence of LLMs has reimagined how guidelines can be integrated into the moderation process. LLMs build their knowledge through extensive pretraining on web-based text, leveraging the transformer architecture \cite{Vaswani2017}. This process employs self-supervised learning, where models predict subsequent words in incomplete sentences, enabling them to associate words and phrases with their typical contexts. As a result, LLMs acquire an internal knowledge of language that enables them to generate coherent and contextually relevant responses across diverse scenarios. Interaction with LLMs relies on prompts—input text that guides the model’s response. LLMs encode the prompt into a high-dimensional vector space, preserving semantic relationships between words and phrases. These representations inform the model’s output, drawing on patterns learned during pretraining. 

\begin{figure}[h]
  \centering
  \includegraphics[width=.7\linewidth]{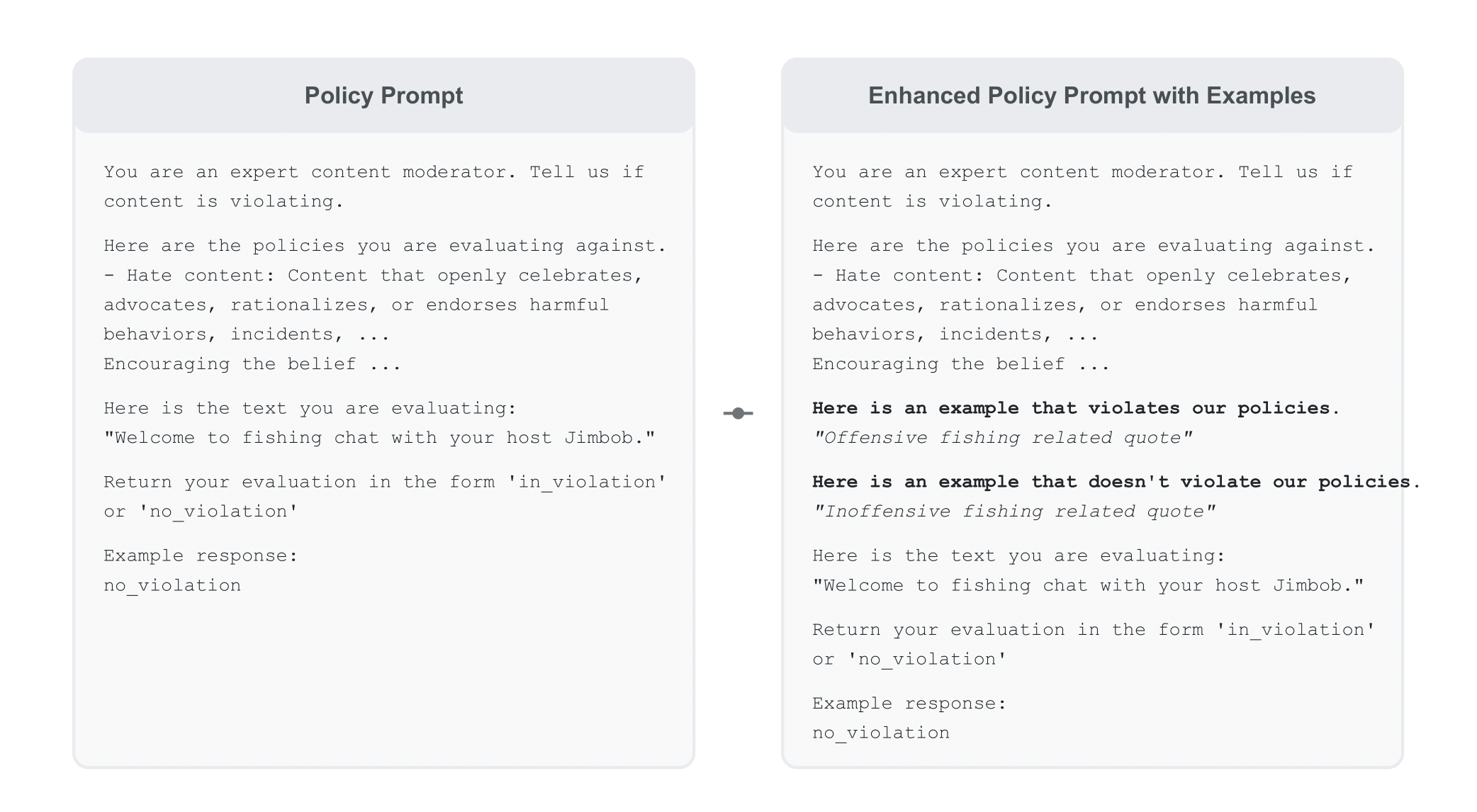}
  \caption{Example prompts demonstrating two approaches for algorithmic content moderation using policy guidelines. \textbf{Left}: A basic prompt where the policy text is provided to the model for direct review of content. \textbf{Right}: An enhanced prompt, following the `in-context learning' technique, that includes both the policy text and specific examples of violative and non-violative content, aiding the model in contextualizing its decisions. }
  \Description{}
  \label{fig:prompts}
\end{figure}

The quality of an LLM’s response is profoundly influenced by the prompt. A well-crafted prompt ensures that the model’s output aligns with the intended outcome, highlighting the prompt as a pivotal element in effective LLM interaction. Factors such as phrasing, specificity, and context within the prompt play a critical role in shaping the response. As a result, \emph{prompt engineering} has emerged as a systematic practice to extract knowledge from LLMs. This involves designing precise, task-specific instructions, that guide models toward accurate, relevant, and coherent outputs without retraining or altering the model’s parameters (weights)~\cite{anthropic2024prompt}.  Techniques range from simple instruction-based prompts to advanced methods like in-context learning, where illustrative examples are included in the prompt to steer the model towards the desired output~\cite{schulhoff2024, sahoo2024systematicsurveypromptengineering} (see Figure \ref{fig:prompts}). Other sophisticated techniques, such as Chain-of-Thought prompting (CoT), encourage the model to break down complex reasoning into step-by-step deductions~\cite{wei2022}, while Reason and Act (ReAct) prompts guide the model to interact with external tools or information sources to enhance its responses~\cite{yao2023react}.

In content moderation, prompts enable quick adaptation of LLMs to changing policies or criteria, offering a flexible approach that avoids the need for retraining (see Figure \ref{fig:policy_prompt_setup}). By simply re-engineering the prompt with updated policy guidelines, LLMs can swiftly respond to evolving societal norms, emerging risks, or platform-specific needs. In this setting, policy guidelines become an integral part of the prompt, allowing for seamless updates and adjustments to content moderation practices. 

LLMs in content moderation currently operate in hybrid setups where human moderators collaborate with and oversee LLM-driven enforcement, but they have the potential to transform content moderation if given greater autonomy. To ensure this does not lead to harm, we need research to understand the complexities of this future. As organizations adopt this paradigm, they must navigate challenges spanning  technical, sociotechnical, organisational, and governance dimensions as we show in Table \ref{tab:area_challenges}. \textbf{Technical implementation} challenges emerge from the fundamental change in how policy guidelines are operationalized--moving from trained models based on annotated datasets to dynamic prompts that directly guide LLM behaviour. This shift introduces concerns around accurately converting policy to prompts and the sensitivity of LLMs to variations in prompt structure and formatting. 
At the \textbf{sociotechnical} level, challenges stem from the limitations of LLMs, which can constrain human judgment and social adaptability. Embedding policies directly into LLM prompts risks reinforcing technological determinism, where policies are rigidly interpreted based on the model’s design, and potentially overlooking the nuanced interplay between technical systems and the human contexts in which they operate. 
On an \textbf{organisational} front, this shift demands changes to workflows and expertise. Traditional content moderation relied on distinct roles between policy formulation and operationalisation, where collaboration was often limited. Policy experts defined guidelines, and machine learning teams operationalised them through annotated datasets. In contrast, policy-as-prompt approaches necessitate hybrid expertise and foster closer collaboration between policy experts and machine learning practitioners.
Finally, the model \textbf{governance} domain encompasses challenges related to transparency and accountability. Ambiguities in decision attribution, coupled with difficulties in tracking prompt modifications, hinder auditability and complicate responsible deployment.

\section{Technical Implementation Challenges}
\label{sec:technical_implementation_challenges}
\subsection{Converting Policies to Prompts}
\label{sec:technical_pol_to_prompt}
Traditionally, policy writing has been a human-centric endeavour, where subject matter experts meticulously crafted guidelines designed for clarity and human interpretation. These guidelines were refined through iterative feedback from human reviewers, including annotators, moderators, and domain experts, who brought contextual understanding, nuanced judgment, and collective experience to the process. Ambiguities were addressed through deliberation and testing. For instance, annotators would extensively test policies by creating hypothetical scenarios, exploring edge cases, and providing qualitative feedback, enhancing the robustness of the guidelines. The primary metric of success was the policy's clarity and actionability for human practitioners.

In the policy-as-prompt paradigm, the policy prompt becomes a distinct artifact. Policy writing now serves a dual purpose: it must remain accessible to human understanding, while also be supplemented with representations optimized for machine interpretation. This requires translating a policy's intent and rules into a format that is optimised for machine processing while faithfully preserving its meaning. However, verifying whether a prompt accurately conveys the policy's intent to the model is challenging. Such verification often relies on other signals, such as comparing LLM moderation decisions against human-labelled ground truth or expert judgments, to assess whether the model applies the policy as intended. 

In traditional supervised learning approach, policies were operationalized through labelled training data, embedding intent within the dataset. While this approach had its own challenges--such as human annotators occasionally misunderstanding policies or models generalized imperfectly--it benefited from redundancy in the dataset, which allowed for statistical smoothing through the sheer volume of labelled examples. In contrast, policy-as-prompt encodes intent directly into the prompt, requiring the  prompt to fully capture the nuances of the policy. Unlike human-centric policy development and enforcement, which relies on human interpretation, writing for machines relies on machine interpretation.
The internal workings of LLMs remain largely opaque, leaving prompt engineering an experimental process guided by trial and error~\cite{bommasani2022opportunitiesrisksfoundationmodels}. This challenge is further compounded by the lack of shared contextual understanding between humans and machines; LLMs often misinterpret nuanced or implicit cues that are easily understood by human readers~\cite{blaise2022, amirizaniani2024llmsexhibithumanlikereasoning}. 

Additionally, as discussed in Section \ref{sec:policy_as_parameter}, the process is highly sensitive to formatting and phrasing. Seemingly minor changes--such as using bullet points instead of paragraphs or altering punctuation--can significantly influence model outputs. Lastly, unlike traditional approaches, there is no intermediary dataset to more exhaustively mitigate these sensitivities, making the translation of policies into prompts, a non-trivial and error-prone task.

%



\subsection{Prompt Structure and Format Sensitivity}
\label{sec:policy_as_parameter} 
In LLM-based content moderation systems, the structure and format of policy guidelines embedded within prompts are critical factors influencing performance.  The presentation and organisation of policy text can directly affect moderation outcomes. Recent empirical studies highlight the significant, yet often overlooked, impact of prompt structure on LLM performance. For instance, \citet{levy-etal-2024-task} found that increasing input length can negatively impact performance, with significant drops occurring well before reaching the models’ maximum input capacity.
Similarly, \cite{liu-etal-2024-lost} showed that performance drops significantly when relevant information is repositioned, highlighting a lack of robustness in long-context processing.
\citet{he2024} demonstrated  that seemingly minor formatting choices--such as using bullet points instead of paragraphs--can substantially influence model performance across tasks. Perhaps most concerning, \citet{sclar2024quantifying} showed that LLMs exhibit unexpected sensitivities to superficial formatting elements like capitalisation and whitespacing, challenging common assumptions about their robustness. 

The precise details of why certain prompt structures work better than others is still an active area of research in AI. Related works have observed that these sensitivities likely emerge from fundamental aspects of LLM architecture and pretraining. The models process text through self-attention mechanisms that compute relationships between all tokens in the input sequence \cite{Vaswani2017}. While theoretically capable of handling arbitrary input structures, the effectiveness of these attention patterns is heavily influenced by how information is organized in the prompt. \citet{zhang2024attentioninstructionamplifyingattention} and \citet{li2024concentrateattentiondomaingeneralizableprompt} demonstrated that different prompt structures lead to distinctly different attention patterns, affecting how information flows through the model's layers. Additionally, \citet{kazemnejad2023the} showed that the positional encoding schemes used in transformer architectures can create inherent biases in how models process information at different positions in the sequence, potentially explaining the observed sensitivity to information positioning.

These findings underscore that the structural choices in presenting information to LLMs are far from superficial. Instead, they represent fundamental parameters that can significantly impact model behavior and reliability. This sensitivity has critical implications for applications like content moderation, where consistent and reliable performance is essential to ensure fairness, accuracy, and trust in automated decisions. Further, this structural sensitivity parallels broader concerns in machine learning regarding how seemingly minor design choices can produce significant downstream effects. Recent research shows that variations in machine learning pipeline components, even when they do not substantially affect overall accuracy metrics, can lead to distinct error patterns or divergent outcomes on specific data subsets. This phenomenon, known as~\textit{predictive multiplicity}, has been observed across various classification tasks and raises particular concerns in high-stakes applications like content moderation~\cite{gomez2024, simson2024}. Similarly, in the policy-as-prompt setting, variations in prompt formatted may result in inconsistent moderation decisions on specific pieces of content, even when overall performance metrics appear unchanged.

\subsubsection{Demonstration}
\label{sec:prompt_structure_format}
To support our discussion, we empirically investigate how the structure, style and the format of policy prompts influence LLM-based content moderation performance. We also further investigate whether predictive multiplicity emerges in this scenario.

\paragraph{Dataset} We collaborated with moderators from a large-scale music platform to create a dataset of 
$2,115$ text descriptions labelled across sixteen different moderation categories. We report in this study six categories, which are widely used in industry for content moderation~\cite{Singhal_2023}: `not violating', `Harm', `Self Harm', `Hate', `Sexual Content' and `jailbreak'. The dataset comprises user-generated requests for playlist creation, offering real-world data for studying moderation. Each entry was labelled as not violating or falling under one specific policy category from the list of considered ones.

\paragraph{Setup}
Our experimental architecture follows the policy-as-prompt framework depicted in Figure~\ref{fig:policy_prompt_setup}.  For the LLM component, we use GPT-4o-mini~\cite{openai2024gpt4o-mini}, for its balance between performance and cost while maintaining characteristics representative of industry-standard models. 
The model's \texttt{{\{\{ prompt \}\}}} included a dictionary with pairs of policy label and policy descriptions for each of the categories considered, 
alongside the preamble instruction: `\texttt{You are a helpful content moderation assistant. You categorize user input into one of these categories depending on whether the input violates the category}'. We use constrained decoding to prevent the model from deviating from the set of considered categories. While this architecture reflects common industry practices, our specific implementation choices--such as model selection, prompt design, and policy structures--were tailored for research purposes and are not representative of production systems.

\begin{figure}[h]
  \centering
  \begin{subfigure}[t]{0.48\linewidth}
    \centering
    \includegraphics[width=\linewidth]{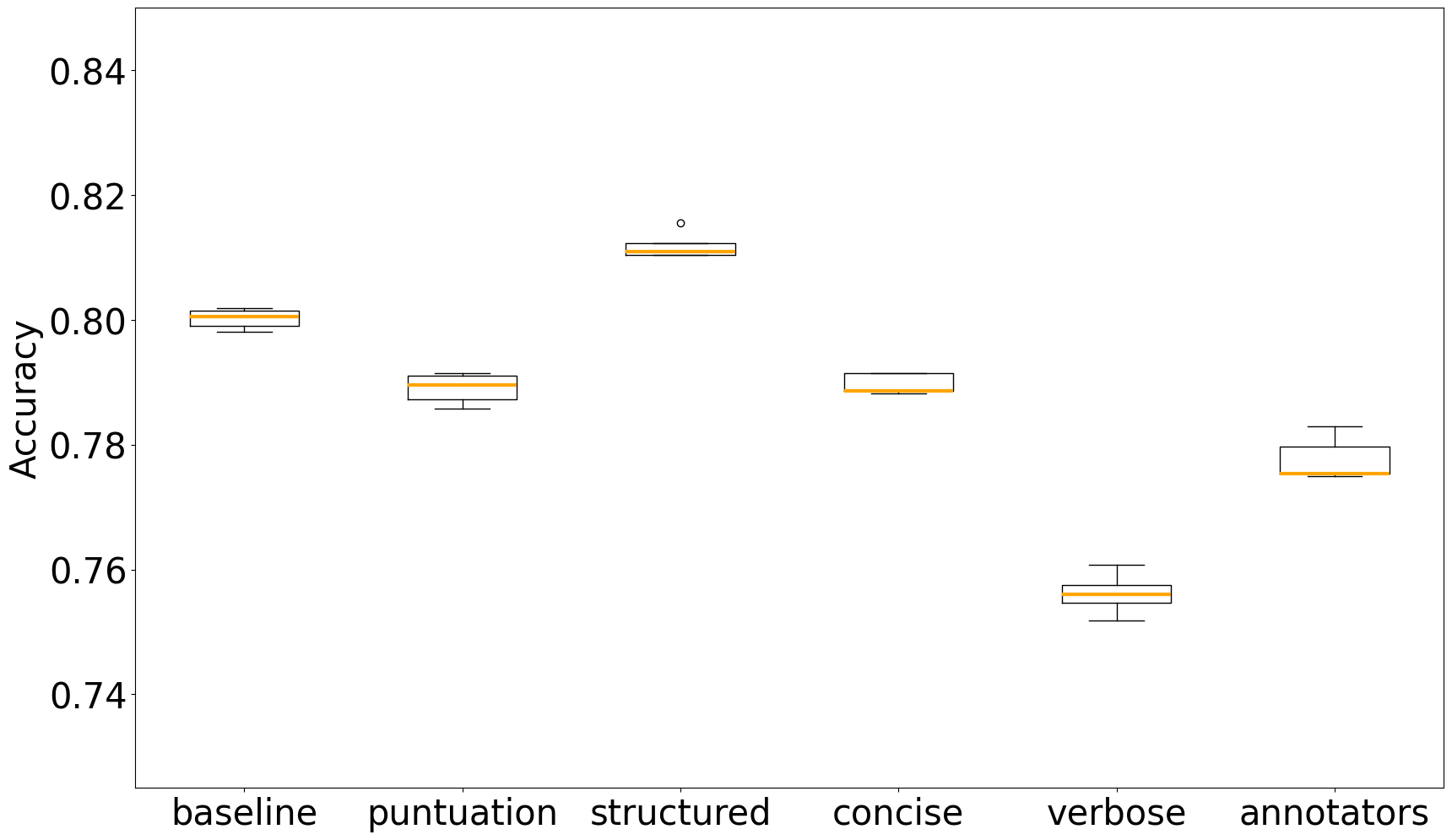}
    \caption{}
    \Description{}
    \label{fig:pred_multi_a}
  \end{subfigure}
  \hfill
  \begin{subfigure}[t]{0.48\linewidth}
    \centering
    \includegraphics[width=\linewidth]{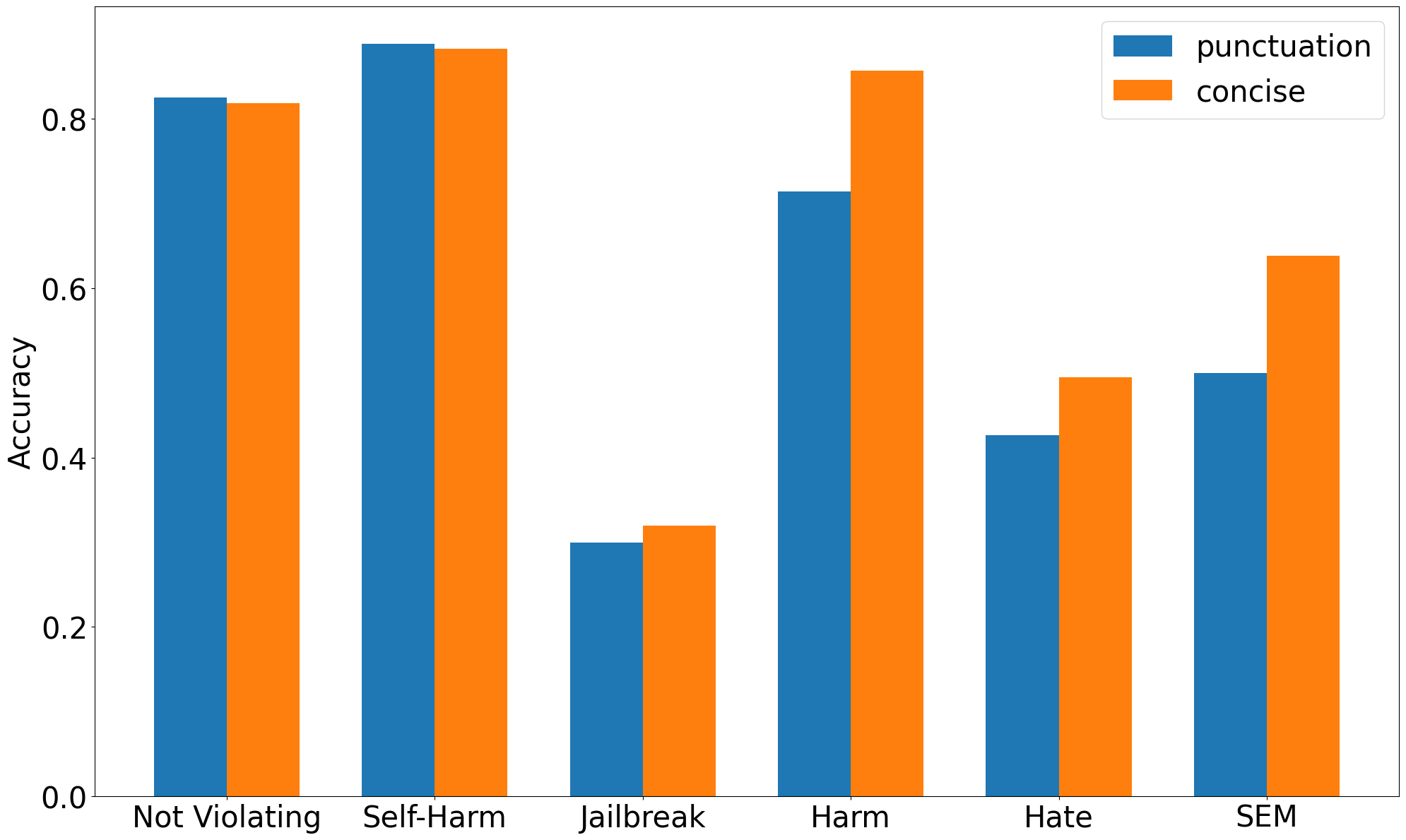}
    \caption{}
    \Description{}
    \label{fig:pred_multi_b}
  \end{subfigure}
  \caption{Effect of prompt design on accuracy: (a) Variation in accuracy across prompt types, averaging over five runs for each type and, (b) Performance differences across policy categories for the `Punctuation' and `Concise' prompt types.}
  \label{fig:pred_multi}
\end{figure}

\begin{figure}[htbp]
    \centering
    \includegraphics[width=.6\linewidth, trim=0 0 0 0, clip]{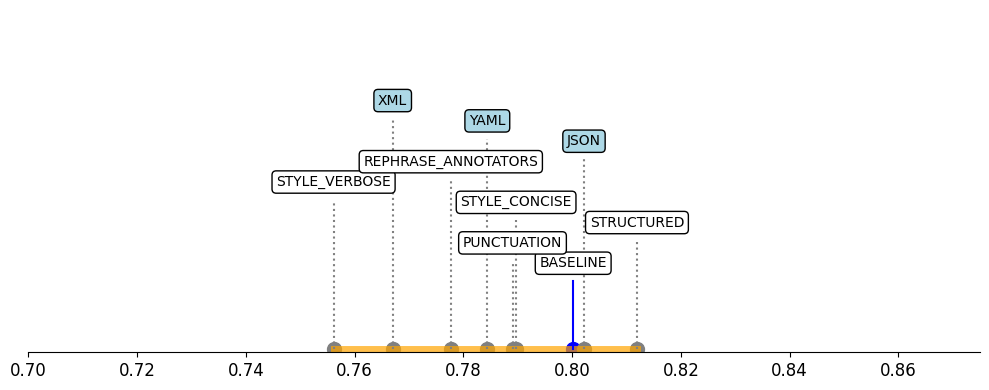} 
    \caption{Performance spread (accuracy) for modifications in the format in which the policy is plugged into the prompt. `Baseline' refers to the plain text format.}
    \Description{}
    \label{fig:policy_format}
\end{figure}

First, we examine how subtle variations in prompt structure and style influence model performance. To investigate this, in Figure~\ref{fig:pred_multi} we focus on cases when semantic differences between templates and the original policy text (the \textit{Baseline}) are minimal. The variations include: \textit{Punctuated}, altered punctuation to assess sensitivity to grammatical changes; \textit{Structured}, used bullet points and sections to enhance clarity and facilitate programmatic processing; \textit{Concise}, streamlined phrasing to test the impact of brevity; \textit{Verbose},  detailed and elaborate descriptions to examine the effect of information density; and \textit{Annotator-Friendly}, tailored for human comprehension, emphasising simplicity and clarity.  Detailed examples of these prompt variations are provided in Figure~\ref{fig:prompt-templates} in Appendix Section~\ref{sec:exp}. 
We further explore how differences in policy prompt formatting--such as plain text, XML, YAML and JSON--affect model performance in Figure~\ref{fig:policy_format}.

Figure~\ref{fig:pred_multi_a} reveals  variability in accuracy across different prompt types, despite the fact that all these variations include similar information. Structured prompts demonstrated the highest accuracy, showing model's preference towards clarity and organisation. In contrast, verbose prompts performed the worst, indicating that excessive detail may hinder the model's understanding.
``Punctuation'' and ``Concise'' prompts showed comparable performance, with average accuracies of 
0.7890 and 0.7897, respectively (\textit{p-value} = 0.631). Figure~\ref{fig:pred_multi_b} compares accuracy across policy categories for the two prompt types, revealing evidence of predictive multiplicity; despite similar overall performance, seemingly identical prompts produced different predictions for the same samples.  The results in Figure \ref{fig:policy_format} confirm that variations in policy formats lead to differences in performance..

\begin{figure}[htbp]
  \centering
  \begin{subfigure}[t]{0.47\linewidth}
    \centering
    \includegraphics[width=0.94\linewidth]{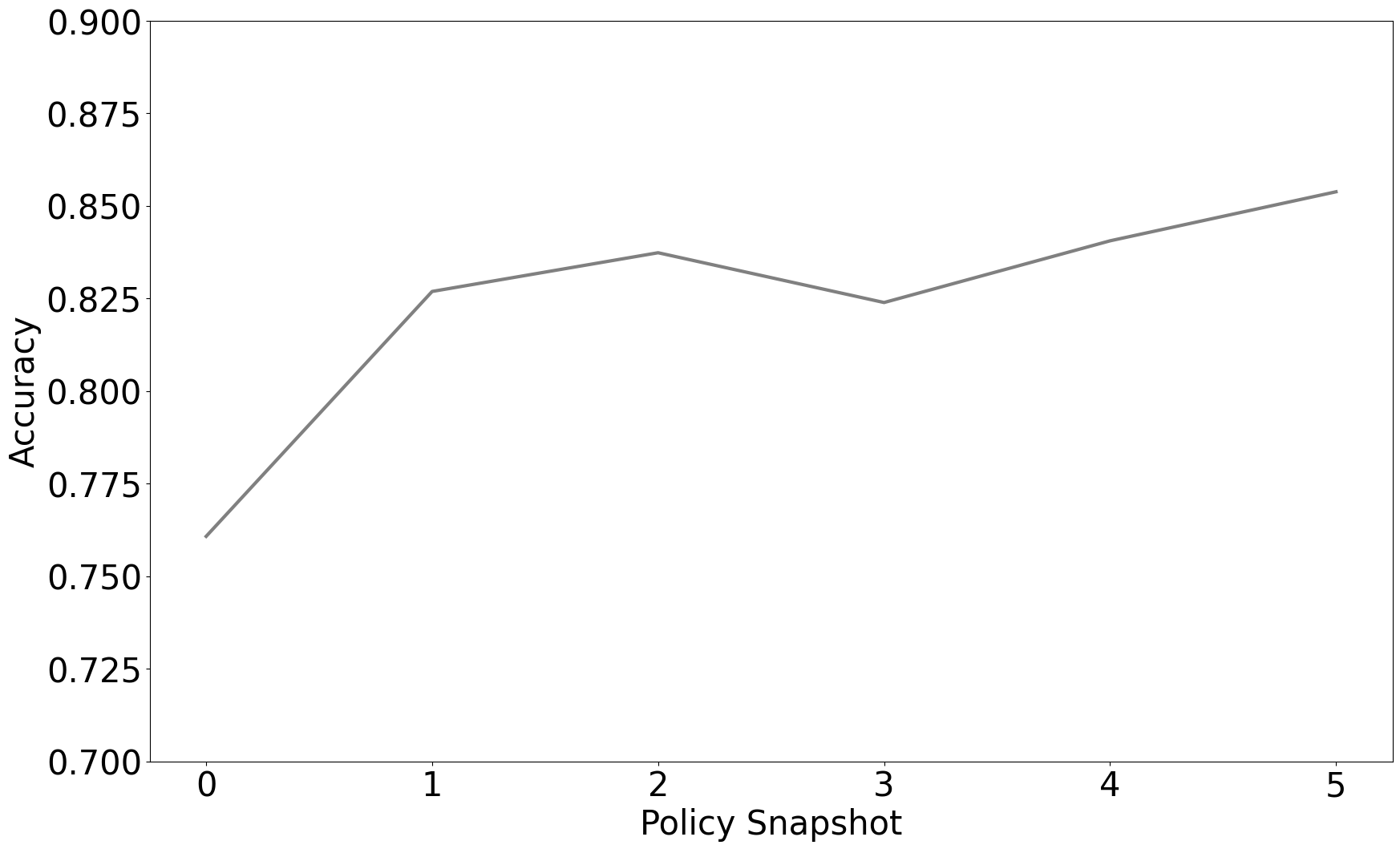}
    \caption{Overall accuracy scores across each temporal snapshot}
    \Description{}
    \label{fig:prompt_temporal_a}
  \end{subfigure}
  \hspace{0.2in}
  \begin{subfigure}[t]{0.47\linewidth}
    \centering
    \includegraphics[width=0.94\linewidth]{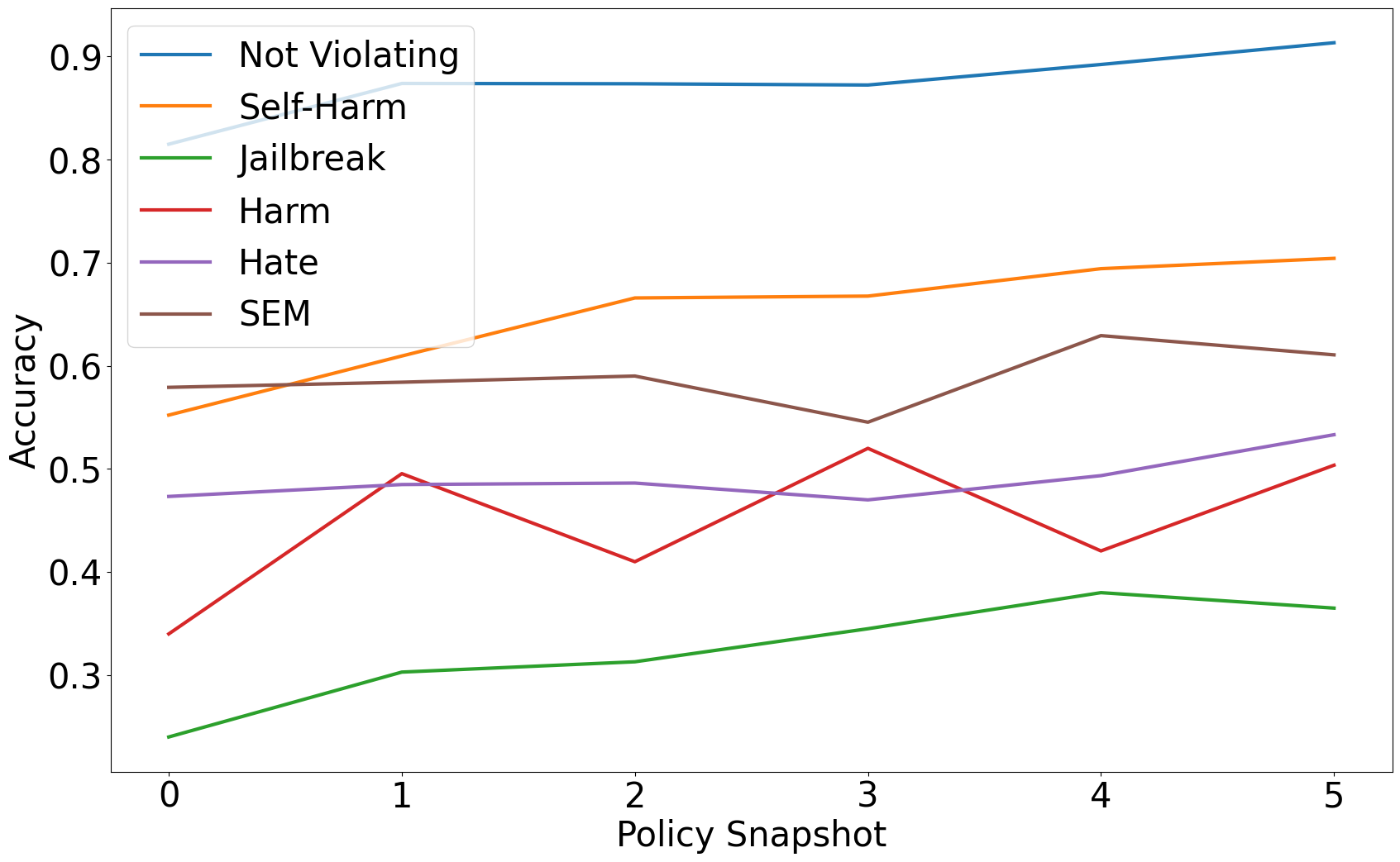}
    \caption{Per-category accuracy scores across each temporal snapshot}
    \Description{}
    \label{fig:prompt_temporal_b}
  \end{subfigure}
  \caption{Analysing both overall and per-category performance of the model when injecting different temporal snapshots of the policies}
  \label{fig:prompt_temporal_all}
\end{figure}

Next, we explore how increasing the amount of policy text in the prompt impacts model performance. To do this, we incrementally add information in the policy descriptions, such as extra facts and examples. This process resulted in sequential snapshots of policies with varying levels of information, with which we analysed for model performance, as illustrated in Figure~\ref{fig:prompt_temporal_a}. This approach mirrors the evolution of policy guidelines on digital platforms, where trends and new boundaries emerge over time. On average, model performance across all policy categories improves as the policy prompt becomes more detailed (greater semantic coverage). However, when examining the model's performance for individual policy categories at each snapshot (Figure \ref{fig:prompt_temporal_b}), we observe variability, suggesting that different categories are influenced to varying degrees by the increase in prompt information.




\section{Sociotechnical Challenge: Technological Determinism in Policy Formation}
\label{sec:sociotechnical_challenges}

Another challenge we identify in the policy-as-prompt approach is that of technological determinism. Technology significantly influences societal values and structures. The concept of technological determinism posits that technology, along with its design and inherent capabilities, can shape social structures, cultural norms, and even political systems~\cite{winner1980artifacts}. Algorithmic systems, particularly those used in content moderation, have long been susceptible to technological determinism due to limitations such as model bias and undertraining \cite{Gorwa2020}. 
Traditionally, human oversight in content moderation served as a vital feedback loop. Human annotators played a key role in shaping training data, ensuring that machine learning models were informed by social contexts and values. This interaction mitigated the influence of technology, enabling the development of nuanced rules that accounted for context and intent, incorporated ethical considerations, and provided mechanisms for human appeals to challenge automated decisions~\cite{roberts2019}.

The advent of LLMs, particularly proprietary and opaque third-party models, is now shifting this balance. By enabling the direct use of policy guidelines in moderation workflows, thus bypassing human annotation, LLMs amplify the risks of technological determinism. As discussed in Section~\ref{sec:technical_implementation_challenges}, LLMs excel at processing structured, algorithmic-friendly guidelines. However, this capability may inadvertently push experts to prioritize machine clarity over the nuance and flexibility required for context-dependent rules. Structured formulations might be favoured even when they do not fully align with underlying policy objectives, creating an environment where complex social issues are reduced to simplistic, binary judgments. This echoes Lessig's ``code as law'' principle, where technical architecture becomes a primary regulator of online behaviour~\cite{lessig1999code}. Similarly, \citet{yeung2019algorithmic} describes ``algorithmic regulation'' as the process by which designing and deploying algorithms transforms into a form of governance, shaping both individual and collective behaviour. Finally, \citet{vandijck2013culture} further argues that platform governance is increasingly dictated by technical affordances rather than democratic deliberation. Consequently, there is a growing risk that content moderation will be driven primarily by what LLMs can efficiently process, rather than by what best serves the diverse needs of online communities.


A broader concern is that LLM capabilities could become the dominant force  shaping policy decisions, overshadowing thoughtful considerations of platform values, user needs, and societal impact. The difficulty of encoding complex ethical considerations into LLM-interpretable rules may push platforms towards simpler, more easily enforceable policies, resulting in the homogenisation of content moderation practices. If LLMs struggle with highly nuanced or context-sensitive rules, platforms may opt for standardized policies to ensure consistent enforcement. This trend recalls the Television Code era of the 1950s and 1960s, when standardized broadcast guidelines homogenized content across local stations, forcing diverse communities to conform to mainstream cultural norms rather than addressing their unique need~\cite{barnouw1970image, macdonald1990one}. Similarly, LLM-driven content moderation risks prioritizing technological efficiency over the diverse and context-specific needs of global online communities, potentially undermining efforts to serve the varied requirements of different cultures and contexts.
\section{Organisational Challenge: Converging Roles, Policy Authors and ML Practitioners}
\label{sec:organisational_practice_challenges}

The integration of promptable LLMs into content moderation workflows has introduced a significant  shift in roles and skill sets for both policy experts and machine learning practitioners. As LLMs are used to operationalise guidelines, policy authors are increasingly required to broaden their expertise beyond traditional policy development to include a working knowledge of machine learning principles, such as prompt engineering and the nuances of algorithmic behaviour, including bias, accuracy, and explainability~\cite{bommasani2022opportunitiesrisksfoundationmodels}. This expanded skill set enables policy experts to anticipate potential issues, assess LLM performance, and contribute to the creation of more robust and fair moderation systems. Conversely, machine learning practitioners are stepping into domains traditionally governed by policy experts. By designing and refining the prompts that guide LLMs, they play a pivotal role in shaping how policies are interpreted and enforced. However, without formal training in policy development or regulatory frameworks, their contributions may inadvertently introduce systemic biases or misinterpretations that diverge from the policy's original intent and potentially overlook critical ethical considerations. 

These emerging workflows are not simply technical translations of policy but complex ``contested spaces'' where different professional epistemologies and interpretive frameworks converge~\cite{green2021}. This intersection highlights the growing need for cross-disciplinary collaboration, where machine learning practitioners develop a deeper understanding of policy nuances, and policy experts acquire foundational machine learning literacy. In this evolving paradigm, content moderation would no longer rely on distinct, sequential contributions from policy experts and machine learning  practitioners. Instead, it would require hybrid expertise and continuous collaboration.

A notable example of this shift is the adoption of red teaming in LLM evaluation~\cite{perez-etal-2022-red, ganguli2022redteaminglanguagemodels}. Traditional evaluation methods, which rely on bespoke testing procedures, often fall short in capturing the full spectrum of behaviours and risks associated with increasingly autonomous and capable LLMs. Red teaming addresses this challenge by borrowing concepts from security practices, where deliberate attack strategies are employed to stress-test systems and uncover vulnerabilities. This approach requires not only technical expertise to understand and exploit the model's intricacies but also insights from fields such as social science, ethics, and policy to identify risks beyond technical performance, such as misuse or harmful societal impacts.

This shift underscores the need for hybrid knowledge systems that blend technical, ethical, and policy expertise. In the future, content moderation may give rise to roles such as ``AI policy translators''--professionals skilled in bridging the gap between technical and policy teams. These individuals would play a pivotal role in ensuring that automated systems align with policy goals while leveraging the capabilities of LLMs. By fostering cross-disciplinary collaboration, they would contribute to the development of more robust, adaptable, and ethically sound moderation practices.

\section{Governance Challenges}
\label{sec:governance_challenges}

The rapid adoption of LLMs in automated decision-making systems has intensified discussions around model governance, focusing on issues such as transparency, accountability, and fairness in the deployment of AI-driven systems \cite{gebru2021datasheets, oecd2019, europeancommission2020}.
Rather than revisiting existing discussions on governance for automated decision-making systems, we specifically examine the implications unique to the policy-as-prompt configuration.


Decision attribution poses an significant accountability challenge in the policy-as-prompt setup. The lack of clear system boundaries create confusion when unwanted outcomes occur, making it difficult to  pinpoint their exact cause--whether they stem from the policy language in the prompt, the LLM's interpretation of that language, or the complex interaction between the two. This ambiguity directly undermines auditability – reconstructing the decision-making process and documenting the influence of various components becomes incredibly difficult. The involvement of multiple stakeholders, including policy writers, prompt engineers, and model providers, further complicates attribution by obscuring how their contributions influence these outcomes~\cite{diakopoulos2016}. 

A further challenge is determining which prompt adjustments warrant formal documentation. As discussed in Section \ref{sec:technical_implementation_challenges}, changes to enforcement can be made by modifying prompts, often with the semantic meaning of the underlying policies remaining unchanged. However, the sensitivity to prompt structure can lead to significant shifts in model behaviour, making it difficult to track how enforcement evolves over time. If every minor change requires extensive documentation, it adds unnecessary overhead. Yet, if changes are not sufficiently documented, it becomes increasingly difficult to monitor and understand the impact of these modifications on model performance. 

\section{Recommendations} \label{sec:mitigations}
In this section, we build on the systematic breakdown of challenges presented earlier (see Table \ref{tab:area_challenges}) to propose strategic directions for addressing them. These strategies are informed by the insights gained from analysing the technical, sociotechnical, organisational and governance-related aspects of the challenges. Rather than offering a definitive set of solutions, this section emphasizes high-level, actionable pathways that can guide future research and practical efforts. By framing these strategies within the context of our earlier analysis, we aim to provide a foundational perspective for tackling the complexities associated with this technology.

\subsection{Enhanced Evaluation }
\label{sec:enhanced_eval}
A key mitigation strategy across several challenges is to do rigorous evaluation of the policy-as-prompt implementation. This evaluation must extend beyond traditional accuracy metrics to assess performance across critical dimensions. 

To address \textit{technical} sensitivity to prompt structure and formatting, the evaluation process should include comprehensive sensitivity analysis during the prompt development phase. This would involve studying the impact of formatting, phrasing, and structural changes on model outputs. Stress testing with diverse policy prompts ensures that small adjustments, such as punctuation or formatting changes, do not unintentionally lead to misinterpretation by the model. \citet{sclar2024quantifying} recommend reporting performance accross a range of plausible prompt formats and styles rather that focusing solely on the performance of a single prompt format. Sensitivity analysis should also address subtle inconsistencies, such as those revealed by predictive multiplicity. For instance, Rashomon sets can identify cases where semantically similar policy phrasing elicit divergent model responses,  such as inconsistent handling of edge cases, despite achieving comparable accuracy metrics~\cite{gomez2024, marx20}. By analysing these variations, developers can identify policy formulations that are both effective and robust. 

At the \textit{sociotechnical} level, evaluation should move beyond conventional accuracy metrics to include measures of societal readiness and adaptability. Metrics like demographic fairness~\cite{mitchell2021, verma2018} can help identifying whether specific prompt formulations create disparities across demographic groups, ensuring equitable policy application. Additionally, case libraries can serve as a valuable tool for addressing potential simplifying tendencies of LLMs. These libraries, containing nuanced, real-world examples of moderation edge cases--such as region-specific cultural references or satire--can test how well a policy-as-prompt system manages societal complexity. Specifically for the prompted guidelines, case libraries become crucial in revealing where binary or reductive prompt designs break down when confronted with complex real-world scenarios. Insights from this edge case analysis can guide iterative refinements of prompts, enabling policy makers and practitioners to design policy-as-prompt configurations that balance technical clarity with societal needs.

\subsection{Collaborative Prompt Engineering}
In Section \ref{sec:technical_pol_to_prompt} we discussed the complexity of converting policy guidelines to prompt, requiring careful attention to the ambiguities that can arise due to the inherent differences between human and machine interpretability of the policy prompt. To minimize machine misinterpretations, strategies must focus on crafting prompts that address multi-faceted aspects of content and encourage diverse perspectives. Already techniques like chain-of-thought reasoning \cite{wei2022} have shown promise in crafting prompts that encourage step-by-step reasoning, allowing models to consider multiple facets of a problem before arriving at a conclusion. 

Techniques like meta-prompting can dynamically integrate contextual ethical reasoning \cite{suzgun2024metapromptingenhancinglanguagemodels}, and multi-persona prompting \cite{wang-etal-2024-unleashing} can embed varied policy perspectives, enabling models to better navigate nuanced contexts.
Additionally, fostering collaborative environments where LLMs—trained on diverse data and architectures—contribute to policy interpretation helps mitigate biases and refine prompt designs. These LLMs can collaboratively suggest rewrites and identify gaps creating a feedback loop for continuous improvement \cite{liang2024cmatmultiagentcollaborationtuning}. 
By treating prompt writing as an iterative, collaborative process of continuous refinement, practitioners can develop more nuanced and adaptive approaches to capturing the complex contextual understanding required in content moderation.
\subsection{Traceability of Prompt Modifications}


At the model \textit{governance } level accountability concerns can be addressed by implementing mechanisms that enhance the transparency and traceability of prompt modifications. A ``prompt genealogy'' could record changes to prompt structures, formatting, and phrasing, along with the rationale behind each modification. Similar to version control systems used in software development~\cite{spinellis2005version} and drawing inspiration from tools like DVC (Data Version Control)~\cite{dvc2020} and Pachyderm~\cite{pachyderm2021}, which provide versioning and lineage tracking for datasets and machine learning pipelines, a prompt genealogy would extend these principles to prompt engineering. 
For instance, practitioners could generate contextual logs or metadata documenting key aspects of the system's behaviour, such as the inputs provided, the policy components referenced in the prompts, and the resulting outputs. While this approach does not reveal the internal workings of how the LLM processes or interprets prompts internally--a subject of ongoing research~\cite{singh2024rethinking}--it offers a structured way to analyse how prompt changes or input variations  correlate with differences in outcomes.

This transparent record would also provide organisations with a comprehensible audit trail, supporting accountability requirements and reproducibility.  It would also allow organisations to demonstrate alignment with intended policy goals and, where necessary, facilitate external reviews or assessments. Finally, in addition to this tool being useful for auditing, it can complement the sensitivity analysis idea mentioned in Section \ref{sec:enhanced_eval},  by providing data for offline tests.


 %

\subsection{Bridging Organisational Silos}
As discussed in Section~\ref{sec:organisational_practice_challenges}, integrating policy authors and machine learning practitioners into a cohesive workflow is essential. However, the transition may encounter practical challenges, similar to historical precedents in other technological paradigm shifts. For example, in the early 2000s the rise of data science in business required close collaboration between domain experts and data specialists~\cite{davenport2012data}. Similarly, the emergence of bioinformatics in the late 20th century necessitated collaboration between biologists and computer scientists before the field gave rise to specialized bioinformaticians~\cite{mitra2022}.

In the short-term, structured collaboration frameworks should be prioritized. These framework  can include regular joint working sessions, shared documentation practices, and established feedback loops, ensuring that policy and machine learning teams work in tandem rather than sequentially. While the long-term goal may be the development of unified roles, these interim measures focus on creating better interfaces between existing specialist teams. This approach enables organisations to maintain effective content moderation while gradually building toward more integrated expertise.

\section{Conclusions}
\label{sec:conclusion}


In this paper, we explored the complexities of the policy-as-prompt paradigm in content moderation through four dimensions: technical, sociotechnical, organizational, and governance. By presenting a structured overview of the challenges practitioners (would) face when transitioning from ``traditional'' machine learning or rule-based systems to the policy-as-prompt paradigm, we aim to support decision-making and underscore the importance of holistically considering the interplay between the technical capabilities of large language models, social and organizational contexts, and governance frameworks. Additionally, we propose potential strategies to mitigate some of these challenges, while acknowledging that many open questions remain regarding the implementation of effective content moderation systems under this paradigm.

Our discussion highlights challenges across these four dimensions without claiming to be exhaustive, and readers may identify additional issues beyond those covered in this paper. Furthermore, while we emphasize  the transformative potential of the policy-as-prompt paradigm, it is important to note that such systems are not currently implemented in a fully autonomous manner. Instead, they typically operate within hybrid setups, where human moderators work alongside and oversee LLM-driven enforcement mechanisms. However, even within this hybrid framework, the increasing adoption of the policy-as-prompt configuration introduces considerable complexities and potential risks that warrant further investigation. Addressing these challenges proactively will be critical to developing more effective hybrid moderation systems that thoughtfully integrate LLM capabilities with human oversight. 

Empirically, we focused on one specific issue--model sensitivity to prompt structure--to demonstrate the brittleness of current LLMs are. We recognize, however, that model performance is influenced by a broader set of parameters beyond prompt structure, including the extent of pre-training, the choice of fine-tuning techniques, model architecture and size, and other factors that were fixed in our experiments. A more comprehensive analysis that examines how these parameters interact would yield deeper insights into their collective impact on model behaviour and reliability in content moderation tasks.


Going forward, as LLM capabilities continue to improve, the policy-as-prompt has the potential to overcome many of its current limitations, enabling organizations to dynamically align enforcement with nuanced and evolving policy objectives.  Sustained exploration of these advancements will be essential in shaping the future of content moderation systems.




\bibliographystyle{ACM-Reference-Format}

\bibliography{main}


\begin{thebibliography}{71}


\ifx \showCODEN    \undefined \def \showCODEN     #1{\unskip}     \fi
\ifx \showDOI      \undefined \def \showDOI       #1{#1}\fi
\ifx \showISBNx    \undefined \def \showISBNx     #1{\unskip}     \fi
\ifx \showISBNxiii \undefined \def \showISBNxiii  #1{\unskip}     \fi
\ifx \showISSN     \undefined \def \showISSN      #1{\unskip}     \fi
\ifx \showLCCN     \undefined \def \showLCCN      #1{\unskip}     \fi
\ifx \shownote     \undefined \def \shownote      #1{#1}          \fi
\ifx \showarticletitle \undefined \def \showarticletitle #1{#1}   \fi
\ifx \showURL      \undefined \def \showURL       {\relax}        \fi
\providecommand\bibfield[2]{#2}
\providecommand\bibinfo[2]{#2}
\providecommand\natexlab[1]{#1}
\providecommand\showeprint[2][]{arXiv:#2}

\bibitem[AlDahoul et~al\mbox{.}(2024)]%
        {aldahoul2024}
\bibfield{author}{\bibinfo{person}{Nouar AlDahoul}, \bibinfo{person}{Myles Joshua~Toledo Tan}, \bibinfo{person}{Harishwar~Reddy Kasireddy}, {and} \bibinfo{person}{Yasir Zaki}.} \bibinfo{year}{2024}\natexlab{}.
\newblock \bibinfo{title}{Advancing Content Moderation: Evaluating Large Language Models for Detecting Sensitive Content Across Text, Images, and Videos}.
\newblock
\showeprint[arxiv]{2411.17123}~[cs.CV]
\urldef\tempurl%
\url{https://arxiv.org/abs/2411.17123}
\showURL{%
\tempurl}


\bibitem[Amirizaniani et~al\mbox{.}(2024)]%
        {amirizaniani2024llmsexhibithumanlikereasoning}
\bibfield{author}{\bibinfo{person}{Maryam Amirizaniani}, \bibinfo{person}{Elias Martin}, \bibinfo{person}{Maryna Sivachenko}, \bibinfo{person}{Afra Mashhadi}, {and} \bibinfo{person}{Chirag Shah}.} \bibinfo{year}{2024}\natexlab{}.
\newblock \bibinfo{title}{Do LLMs Exhibit Human-Like Reasoning? Evaluating Theory of Mind in LLMs for Open-Ended Responses}.
\newblock
\showeprint[arxiv]{2406.05659}~[cs.CL]
\urldef\tempurl%
\url{https://arxiv.org/abs/2406.05659}
\showURL{%
\tempurl}


\bibitem[Anthropic(2023)]%
        {anthropic2023claude}
\bibfield{author}{\bibinfo{person}{Anthropic}.} \bibinfo{year}{2023}\natexlab{}.
\newblock \bibinfo{title}{Introducing Claude}.
\newblock \bibinfo{howpublished}{Available at \texttt{https://www.anthropic.com/index/introducing-claude}}.
\newblock
\newblock
\shownote{Accessed January 9, 2025}.


\bibitem[Anthropic(2024)]%
        {anthropic2024prompt}
\bibfield{author}{\bibinfo{person}{Anthropic}.} \bibinfo{year}{2024}\natexlab{}.
\newblock \bibinfo{title}{Prompt Engineering Overview}.
\newblock \bibinfo{howpublished}{\url{https://docs.anthropic.com/en/docs/build-with-claude/prompt-engineering/overview}}.
\newblock
\newblock
\shownote{Online; accessed 2024}.


\bibitem[{Anthropic}(2025)]%
        {anthropic_content_moderation}
\bibfield{author}{\bibinfo{person}{{Anthropic}}.} \bibinfo{year}{2025}\natexlab{}.
\newblock \bibinfo{title}{{Use Case Guide: Content Moderation with Claude}}.
\newblock
\urldef\tempurl%
\url{https://docs.anthropic.com/en/docs/about-claude/use-case-guides/content-moderation}
\showURL{%
\tempurl}
\newblock
\shownote{Accessed: 2025-01-11}.


\bibitem[Barnouw(1970)]%
        {barnouw1970image}
\bibfield{author}{\bibinfo{person}{Erik Barnouw}.} \bibinfo{year}{1970}\natexlab{}.
\newblock \bibinfo{booktitle}{\emph{The Image Empire: A History of Broadcasting in the United States, Volume III-from 1953}}.
\newblock \bibinfo{publisher}{Oxford University Press}, \bibinfo{address}{New York}.
\newblock
\showISBNx{0195012593}


\bibitem[Belloni(2023)]%
        {belloni2023multilingual}
\bibfield{author}{\bibinfo{person}{Mattia Belloni}.} \bibinfo{year}{2023}\natexlab{}.
\newblock \bibinfo{title}{Multilingual message content moderation at scale}.
\newblock \bibinfo{howpublished}{\url{https://medium.com/bumble-tech/multilingual-message-content-moderation-at-scale-ddd0da1e23ed}}.
\newblock
\newblock
\shownote{Accessed: January 9, 2025}.


\bibitem[Bommasani et~al\mbox{.}(2022)]%
        {bommasani2022opportunitiesrisksfoundationmodels}
\bibfield{author}{\bibinfo{person}{Rishi Bommasani}, \bibinfo{person}{Drew~A. Hudson}, \bibinfo{person}{Ehsan Adeli}, \bibinfo{person}{Russ Altman}, \bibinfo{person}{Simran Arora}, \bibinfo{person}{Sydney von Arx}, \bibinfo{person}{Michael~S. Bernstein}, \bibinfo{person}{Jeannette Bohg}, \bibinfo{person}{Antoine Bosselut}, \bibinfo{person}{Emma Brunskill}, \bibinfo{person}{Erik Brynjolfsson}, \bibinfo{person}{Shyamal Buch}, \bibinfo{person}{Dallas Card}, \bibinfo{person}{Rodrigo Castellon}, \bibinfo{person}{Niladri Chatterji}, \bibinfo{person}{Annie Chen}, \bibinfo{person}{Kathleen Creel}, \bibinfo{person}{Jared~Quincy Davis}, \bibinfo{person}{Dora Demszky}, \bibinfo{person}{Chris Donahue}, \bibinfo{person}{Moussa Doumbouya}, \bibinfo{person}{Esin Durmus}, \bibinfo{person}{Stefano Ermon}, \bibinfo{person}{John Etchemendy}, \bibinfo{person}{Kawin Ethayarajh}, \bibinfo{person}{Li Fei-Fei}, \bibinfo{person}{Chelsea Finn}, \bibinfo{person}{Trevor Gale}, \bibinfo{person}{Lauren Gillespie}, \bibinfo{person}{Karan
  Goel}, \bibinfo{person}{Noah Goodman}, \bibinfo{person}{Shelby Grossman}, \bibinfo{person}{Neel Guha}, \bibinfo{person}{Tatsunori Hashimoto}, \bibinfo{person}{Peter Henderson}, \bibinfo{person}{John Hewitt}, \bibinfo{person}{Daniel~E. Ho}, \bibinfo{person}{Jenny Hong}, \bibinfo{person}{Kyle Hsu}, \bibinfo{person}{Jing Huang}, \bibinfo{person}{Thomas Icard}, \bibinfo{person}{Saahil Jain}, \bibinfo{person}{Dan Jurafsky}, \bibinfo{person}{Pratyusha Kalluri}, \bibinfo{person}{Siddharth Karamcheti}, \bibinfo{person}{Geoff Keeling}, \bibinfo{person}{Fereshte Khani}, \bibinfo{person}{Omar Khattab}, \bibinfo{person}{Pang~Wei Koh}, \bibinfo{person}{Mark Krass}, \bibinfo{person}{Ranjay Krishna}, \bibinfo{person}{Rohith Kuditipudi}, \bibinfo{person}{Ananya Kumar}, \bibinfo{person}{Faisal Ladhak}, \bibinfo{person}{Mina Lee}, \bibinfo{person}{Tony Lee}, \bibinfo{person}{Jure Leskovec}, \bibinfo{person}{Isabelle Levent}, \bibinfo{person}{Xiang~Lisa Li}, \bibinfo{person}{Xuechen Li}, \bibinfo{person}{Tengyu Ma},
  \bibinfo{person}{Ali Malik}, \bibinfo{person}{Christopher~D. Manning}, \bibinfo{person}{Suvir Mirchandani}, \bibinfo{person}{Eric Mitchell}, \bibinfo{person}{Zanele Munyikwa}, \bibinfo{person}{Suraj Nair}, \bibinfo{person}{Avanika Narayan}, \bibinfo{person}{Deepak Narayanan}, \bibinfo{person}{Ben Newman}, \bibinfo{person}{Allen Nie}, \bibinfo{person}{Juan~Carlos Niebles}, \bibinfo{person}{Hamed Nilforoshan}, \bibinfo{person}{Julian Nyarko}, \bibinfo{person}{Giray Ogut}, \bibinfo{person}{Laurel Orr}, \bibinfo{person}{Isabel Papadimitriou}, \bibinfo{person}{Joon~Sung Park}, \bibinfo{person}{Chris Piech}, \bibinfo{person}{Eva Portelance}, \bibinfo{person}{Christopher Potts}, \bibinfo{person}{Aditi Raghunathan}, \bibinfo{person}{Rob Reich}, \bibinfo{person}{Hongyu Ren}, \bibinfo{person}{Frieda Rong}, \bibinfo{person}{Yusuf Roohani}, \bibinfo{person}{Camilo Ruiz}, \bibinfo{person}{Jack Ryan}, \bibinfo{person}{Christopher Ré}, \bibinfo{person}{Dorsa Sadigh}, \bibinfo{person}{Shiori Sagawa},
  \bibinfo{person}{Keshav Santhanam}, \bibinfo{person}{Andy Shih}, \bibinfo{person}{Krishnan Srinivasan}, \bibinfo{person}{Alex Tamkin}, \bibinfo{person}{Rohan Taori}, \bibinfo{person}{Armin~W. Thomas}, \bibinfo{person}{Florian Tramèr}, \bibinfo{person}{Rose~E. Wang}, \bibinfo{person}{William Wang}, \bibinfo{person}{Bohan Wu}, \bibinfo{person}{Jiajun Wu}, \bibinfo{person}{Yuhuai Wu}, \bibinfo{person}{Sang~Michael Xie}, \bibinfo{person}{Michihiro Yasunaga}, \bibinfo{person}{Jiaxuan You}, \bibinfo{person}{Matei Zaharia}, \bibinfo{person}{Michael Zhang}, \bibinfo{person}{Tianyi Zhang}, \bibinfo{person}{Xikun Zhang}, \bibinfo{person}{Yuhui Zhang}, \bibinfo{person}{Lucia Zheng}, \bibinfo{person}{Kaitlyn Zhou}, {and} \bibinfo{person}{Percy Liang}.} \bibinfo{year}{2022}\natexlab{}.
\newblock \bibinfo{title}{On the Opportunities and Risks of Foundation Models}.
\newblock
\showeprint[arxiv]{2108.07258}~[cs.LG]
\urldef\tempurl%
\url{https://arxiv.org/abs/2108.07258}
\showURL{%
\tempurl}


\bibitem[Commission(2020)]%
        {europeancommission2020}
\bibfield{author}{\bibinfo{person}{European Commission}.} \bibinfo{year}{2020}\natexlab{}.
\newblock \bibinfo{title}{White Paper on Artificial Intelligence - A European approach to excellence and trust}.
\newblock \bibinfo{howpublished}{\url{https://ec.europa.eu/info/publications/white-paper-artificial-intelligence-european-approach-excellence-and-trust_en}}.
\newblock


\bibitem[Davenport and Patil(2012)]%
        {davenport2012data}
\bibfield{author}{\bibinfo{person}{Thomas~H. Davenport} {and} \bibinfo{person}{D.J. Patil}.} \bibinfo{year}{2012}\natexlab{}.
\newblock \showarticletitle{Data Scientist: The Sexiest Job of the 21st Century}.
\newblock \bibinfo{journal}{\emph{Harvard Business Review}} \bibinfo{volume}{90}, \bibinfo{number}{10} (\bibinfo{year}{2012}), \bibinfo{pages}{70--76}.
\newblock
\urldef\tempurl%
\url{https://hbr.org/2012/10/data-scientist-the-sexiest-job-of-the-21st-century}
\showURL{%
\tempurl}


\bibitem[Desai et~al\mbox{.}(2024)]%
        {desai2024genaiusersafetysurvey}
\bibfield{author}{\bibinfo{person}{Akshar~Prabhu Desai}, \bibinfo{person}{Tejasvi Ravi}, \bibinfo{person}{Mohammad Luqman}, \bibinfo{person}{Mohit Sharma}, \bibinfo{person}{Nithya Kota}, {and} \bibinfo{person}{Pranjul Yadav}.} \bibinfo{year}{2024}\natexlab{}.
\newblock \bibinfo{title}{Gen-AI for User Safety: A Survey}.
\newblock
\showeprint[arxiv]{2411.06606}~[cs.AI]
\urldef\tempurl%
\url{https://arxiv.org/abs/2411.06606}
\showURL{%
\tempurl}


\bibitem[Devlin et~al\mbox{.}(2019)]%
        {devlin-etal-2019-bert}
\bibfield{author}{\bibinfo{person}{Jacob Devlin}, \bibinfo{person}{Ming-Wei Chang}, \bibinfo{person}{Kenton Lee}, {and} \bibinfo{person}{Kristina Toutanova}.} \bibinfo{year}{2019}\natexlab{}.
\newblock \showarticletitle{{BERT}: Pre-training of Deep Bidirectional Transformers for Language Understanding}. In \bibinfo{booktitle}{\emph{Proceedings of the 2019 Conference of the North {A}merican Chapter of the Association for Computational Linguistics: Human Language Technologies, Volume 1 (Long and Short Papers)}}, \bibfield{editor}{\bibinfo{person}{Jill Burstein}, \bibinfo{person}{Christy Doran}, {and} \bibinfo{person}{Thamar Solorio}} (Eds.). \bibinfo{publisher}{Association for Computational Linguistics}, \bibinfo{address}{Minneapolis, Minnesota}, \bibinfo{pages}{4171--4186}.
\newblock
\urldef\tempurl%
\url{https://doi.org/10.18653/v1/N19-1423}
\showDOI{\tempurl}


\bibitem[Diakopoulos(2016)]%
        {diakopoulos2016}
\bibfield{author}{\bibinfo{person}{Nicholas Diakopoulos}.} \bibinfo{year}{2016}\natexlab{}.
\newblock \showarticletitle{Accountability in algorithmic decision making}.
\newblock \bibinfo{journal}{\emph{Commun. ACM}} \bibinfo{volume}{59}, \bibinfo{number}{2} (\bibinfo{date}{Jan.} \bibinfo{year}{2016}), \bibinfo{pages}{56–62}.
\newblock
\showISSN{0001-0782}
\urldef\tempurl%
\url{https://doi.org/10.1145/2844110}
\showDOI{\tempurl}


\bibitem[Engstrom and Feamster(2017)]%
        {engstrom2017limits}
\bibfield{author}{\bibinfo{person}{Evan Engstrom} {and} \bibinfo{person}{Nick Feamster}.} \bibinfo{year}{2017}\natexlab{}.
\newblock \bibinfo{booktitle}{\emph{The Limits of Filtering: A Look at the Functionality \& Shortcomings of Content Detection Tools}}.
\newblock \bibinfo{type}{{T}echnical {R}eport}. \bibinfo{institution}{Engine}.
\newblock
\urldef\tempurl%
\url{https://www.engine.is/the-limits-of-filtering}
\showURL{%
\tempurl}


\bibitem[Ganguli et~al\mbox{.}(2022)]%
        {ganguli2022redteaminglanguagemodels}
\bibfield{author}{\bibinfo{person}{Deep Ganguli}, \bibinfo{person}{Liane Lovitt}, \bibinfo{person}{Jackson Kernion}, \bibinfo{person}{Amanda Askell}, \bibinfo{person}{Yuntao Bai}, \bibinfo{person}{Saurav Kadavath}, \bibinfo{person}{Ben Mann}, \bibinfo{person}{Ethan Perez}, \bibinfo{person}{Nicholas Schiefer}, \bibinfo{person}{Kamal Ndousse}, \bibinfo{person}{Andy Jones}, \bibinfo{person}{Sam Bowman}, \bibinfo{person}{Anna Chen}, \bibinfo{person}{Tom Conerly}, \bibinfo{person}{Nova DasSarma}, \bibinfo{person}{Dawn Drain}, \bibinfo{person}{Nelson Elhage}, \bibinfo{person}{Sheer El-Showk}, \bibinfo{person}{Stanislav Fort}, \bibinfo{person}{Zac Hatfield-Dodds}, \bibinfo{person}{Tom Henighan}, \bibinfo{person}{Danny Hernandez}, \bibinfo{person}{Tristan Hume}, \bibinfo{person}{Josh Jacobson}, \bibinfo{person}{Scott Johnston}, \bibinfo{person}{Shauna Kravec}, \bibinfo{person}{Catherine Olsson}, \bibinfo{person}{Sam Ringer}, \bibinfo{person}{Eli Tran-Johnson}, \bibinfo{person}{Dario Amodei}, \bibinfo{person}{Tom
  Brown}, \bibinfo{person}{Nicholas Joseph}, \bibinfo{person}{Sam McCandlish}, \bibinfo{person}{Chris Olah}, \bibinfo{person}{Jared Kaplan}, {and} \bibinfo{person}{Jack Clark}.} \bibinfo{year}{2022}\natexlab{}.
\newblock \bibinfo{title}{Red Teaming Language Models to Reduce Harms: Methods, Scaling Behaviors, and Lessons Learned}.
\newblock
\showeprint[arxiv]{2209.07858}~[cs.CL]
\urldef\tempurl%
\url{https://arxiv.org/abs/2209.07858}
\showURL{%
\tempurl}


\bibitem[Gebru et~al\mbox{.}(2021)]%
        {gebru2021datasheets}
\bibfield{author}{\bibinfo{person}{Timnit Gebru}, \bibinfo{person}{Jamie Morgenstern}, \bibinfo{person}{Briana Vecchione}, \bibinfo{person}{Jennifer~Wortman Vaughan}, \bibinfo{person}{Hanna Wallach}, \bibinfo{person}{Hal Daumé~III}, {and} \bibinfo{person}{Kate Crawford}.} \bibinfo{year}{2021}\natexlab{}.
\newblock \showarticletitle{Datasheets for Datasets}.
\newblock \bibinfo{journal}{\emph{Commun. ACM}} \bibinfo{volume}{64}, \bibinfo{number}{12} (\bibinfo{date}{December} \bibinfo{year}{2021}), \bibinfo{pages}{86--92}.
\newblock
\urldef\tempurl%
\url{https://www.microsoft.com/en-us/research/publication/datasheets-for-datasets/}
\showURL{%
\tempurl}


\bibitem[Gillespie(2018)]%
        {Gillespie2018}
\bibfield{author}{\bibinfo{person}{Tarleton Gillespie}.} \bibinfo{year}{2018}\natexlab{}.
\newblock \bibinfo{booktitle}{\emph{Custodians of the Internet: Platforms, Content Moderation, and the Hidden Decisions That Shape Social Media}}.
\newblock 1--288 pages.
\newblock
\showISBNx{9780300235029}
\urldef\tempurl%
\url{https://doi.org/10.12987/9780300235029}
\showDOI{\tempurl}


\bibitem[Gomez et~al\mbox{.}(2024)]%
        {gomez2024}
\bibfield{author}{\bibinfo{person}{Juan~Felipe Gomez}, \bibinfo{person}{Caio Machado}, \bibinfo{person}{Lucas~Monteiro Paes}, {and} \bibinfo{person}{Flavio Calmon}.} \bibinfo{year}{2024}\natexlab{}.
\newblock \showarticletitle{Algorithmic Arbitrariness in Content Moderation}. In \bibinfo{booktitle}{\emph{Proceedings of the 2024 ACM Conference on Fairness, Accountability, and Transparency}} (Rio de Janeiro, Brazil) \emph{(\bibinfo{series}{FAccT '24})}. \bibinfo{publisher}{Association for Computing Machinery}, \bibinfo{address}{New York, NY, USA}, \bibinfo{pages}{2234–2253}.
\newblock
\showISBNx{9798400704505}
\urldef\tempurl%
\url{https://doi.org/10.1145/3630106.3659036}
\showDOI{\tempurl}


\bibitem[Gorwa et~al\mbox{.}(2020)]%
        {Gorwa2020}
\bibfield{author}{\bibinfo{person}{Robert Gorwa}, \bibinfo{person}{Reuben Binns}, {and} \bibinfo{person}{Christian Katzenbach}.} \bibinfo{year}{2020}\natexlab{}.
\newblock \showarticletitle{Algorithmic content moderation: Technical and political challenges in the automation of platform governance}.
\newblock \bibinfo{journal}{\emph{Big Data \& Society}} \bibinfo{volume}{7}, \bibinfo{number}{1} (\bibinfo{year}{2020}), \bibinfo{pages}{2053951719897945}.
\newblock
\urldef\tempurl%
\url{https://doi.org/10.1177/2053951719897945}
\showDOI{\tempurl}
\showeprint{https://doi.org/10.1177/2053951719897945}


\bibitem[Green(2021)]%
        {green2021}
\bibfield{author}{\bibinfo{person}{Ben Green}.} \bibinfo{year}{2021}\natexlab{}.
\newblock \showarticletitle{The Contestation of Tech Ethics: A Sociotechnical Approach to Ethics and Technology in Action}.
\newblock \bibinfo{journal}{\emph{SSRN Electronic Journal}} (\bibinfo{date}{01} \bibinfo{year}{2021}).
\newblock
\urldef\tempurl%
\url{https://doi.org/10.2139/ssrn.3859358}
\showDOI{\tempurl}


\bibitem[He et~al\mbox{.}(2024)]%
        {he2024}
\bibfield{author}{\bibinfo{person}{Jia He}, \bibinfo{person}{Mukund Rungta}, \bibinfo{person}{David Koleczek}, \bibinfo{person}{Arshdeep Sekhon}, \bibinfo{person}{Franklin~X Wang}, {and} \bibinfo{person}{Sadid Hasan}.} \bibinfo{year}{2024}\natexlab{}.
\newblock \bibinfo{title}{Does Prompt Formatting Have Any Impact on LLM Performance?}
\newblock
\showeprint[arxiv]{2411.10541}~[cs.CL]
\urldef\tempurl%
\url{https://arxiv.org/abs/2411.10541}
\showURL{%
\tempurl}


\bibitem[Inan et~al\mbox{.}(2023)]%
        {inan2023llama}
\bibfield{author}{\bibinfo{person}{Hakan Inan}, \bibinfo{person}{Kartikeya Upasani}, \bibinfo{person}{Jianfeng Chi}, \bibinfo{person}{Rashi Rungta}, \bibinfo{person}{Krithika Iyer}, \bibinfo{person}{Yuning Mao}, \bibinfo{person}{Michael Tontchev}, \bibinfo{person}{Qing Hu}, \bibinfo{person}{Brian Fuller}, \bibinfo{person}{Davide Testuggine}, {and} \bibinfo{person}{Madian Khabsa}.} \bibinfo{year}{2023}\natexlab{}.
\newblock \bibinfo{title}{Llama Guard: LLM-based Input-Output Safeguard for Human-AI Conversations}.
\newblock
\showeprint[arxiv]{2312.06674}~[cs.CL]


\bibitem[Iterative.ai(2020)]%
        {dvc2020}
\bibfield{author}{\bibinfo{person}{Iterative.ai}.} \bibinfo{year}{2020}\natexlab{}.
\newblock \bibinfo{title}{Data Version Control (DVC): Open-source Version Control System for Machine Learning Projects}.
\newblock \bibinfo{howpublished}{\url{https://dvc.org}}.
\newblock


\bibitem[Kazemnejad et~al\mbox{.}(2023)]%
        {kazemnejad2023the}
\bibfield{author}{\bibinfo{person}{Amirhossein Kazemnejad}, \bibinfo{person}{Inkit Padhi}, \bibinfo{person}{Karthikeyan Natesan}, \bibinfo{person}{Payel Das}, {and} \bibinfo{person}{Siva Reddy}.} \bibinfo{year}{2023}\natexlab{}.
\newblock \showarticletitle{The Impact of Positional Encoding on Length Generalization in Transformers}. In \bibinfo{booktitle}{\emph{Thirty-seventh Conference on Neural Information Processing Systems}}.
\newblock
\urldef\tempurl%
\url{https://openreview.net/forum?id=Drrl2gcjzl}
\showURL{%
\tempurl}


\bibitem[Kolla et~al\mbox{.}(2024)]%
        {kolla2024}
\bibfield{author}{\bibinfo{person}{Mahi Kolla}, \bibinfo{person}{Siddharth Salunkhe}, \bibinfo{person}{Eshwar Chandrasekharan}, {and} \bibinfo{person}{Koustuv Saha}.} \bibinfo{year}{2024}\natexlab{}.
\newblock \showarticletitle{LLM-Mod: Can Large Language Models Assist Content Moderation?}. In \bibinfo{booktitle}{\emph{CHI 2024 - Extended Abstracts of the 2024 CHI Conference on Human Factors in Computing Sytems}} \emph{(\bibinfo{series}{Conference on Human Factors in Computing Systems - Proceedings})}. \bibinfo{publisher}{Association for Computing Machinery}, \bibinfo{address}{United States}.
\newblock
\urldef\tempurl%
\url{https://doi.org/10.1145/3613905.3650828}
\showDOI{\tempurl}
\newblock
\shownote{Publisher Copyright: {\textcopyright} 2024 Association for Computing Machinery. All rights reserved.; 2024 CHI Conference on Human Factors in Computing Sytems, CHI EA 2024 ; Conference date: 11-05-2024 Through 16-05-2024}.


\bibitem[Kumar et~al\mbox{.}(2024)]%
        {kumar2024}
\bibfield{author}{\bibinfo{person}{Deepak Kumar}, \bibinfo{person}{Yousef Abuhashem}, {and} \bibinfo{person}{Zakir Durumeric}.} \bibinfo{year}{2024}\natexlab{}.
\newblock \showarticletitle{Watch Your Language: Investigating Content Moderation with Large Language Models}.
\newblock \bibinfo{journal}{\emph{Proceedings of the International AAAI Conference on Web and Social Media}}  \bibinfo{volume}{18} (\bibinfo{date}{05} \bibinfo{year}{2024}), \bibinfo{pages}{865--878}.
\newblock
\urldef\tempurl%
\url{https://doi.org/10.1609/icwsm.v18i1.31358}
\showDOI{\tempurl}


\bibitem[Larimore et~al\mbox{.}(2021)]%
        {larimore-etal-2021-reconsidering}
\bibfield{author}{\bibinfo{person}{Savannah Larimore}, \bibinfo{person}{Ian Kennedy}, \bibinfo{person}{Breon Haskett}, {and} \bibinfo{person}{Alina Arseniev-Koehler}.} \bibinfo{year}{2021}\natexlab{}.
\newblock \showarticletitle{Reconsidering Annotator Disagreement about Racist Language: Noise or Signal?}. In \bibinfo{booktitle}{\emph{Proceedings of the Ninth International Workshop on Natural Language Processing for Social Media}}, \bibfield{editor}{\bibinfo{person}{Lun-Wei Ku} {and} \bibinfo{person}{Cheng-Te Li}} (Eds.). \bibinfo{publisher}{Association for Computational Linguistics}, \bibinfo{address}{Online}, \bibinfo{pages}{81--90}.
\newblock
\urldef\tempurl%
\url{https://doi.org/10.18653/v1/2021.socialnlp-1.7}
\showDOI{\tempurl}


\bibitem[Lessig(1999)]%
        {lessig1999code}
\bibfield{author}{\bibinfo{person}{Lawrence Lessig}.} \bibinfo{year}{1999}\natexlab{}.
\newblock \bibinfo{booktitle}{\emph{Code and Other Laws of Cyberspace}}.
\newblock \bibinfo{publisher}{Basic Books}.
\newblock


\bibitem[Levy et~al\mbox{.}(2024)]%
        {levy-etal-2024-task}
\bibfield{author}{\bibinfo{person}{Mosh Levy}, \bibinfo{person}{Alon Jacoby}, {and} \bibinfo{person}{Yoav Goldberg}.} \bibinfo{year}{2024}\natexlab{}.
\newblock \showarticletitle{Same Task, More Tokens: the Impact of Input Length on the Reasoning Performance of Large Language Models}. In \bibinfo{booktitle}{\emph{Proceedings of the 62nd Annual Meeting of the Association for Computational Linguistics (Volume 1: Long Papers)}}, \bibfield{editor}{\bibinfo{person}{Lun-Wei Ku}, \bibinfo{person}{Andre Martins}, {and} \bibinfo{person}{Vivek Srikumar}} (Eds.). \bibinfo{publisher}{Association for Computational Linguistics}, \bibinfo{address}{Bangkok, Thailand}, \bibinfo{pages}{15339--15353}.
\newblock
\urldef\tempurl%
\url{https://doi.org/10.18653/v1/2024.acl-long.818}
\showDOI{\tempurl}


\bibitem[Li et~al\mbox{.}(2024)]%
        {li2024concentrateattentiondomaingeneralizableprompt}
\bibfield{author}{\bibinfo{person}{Chengzhengxu Li}, \bibinfo{person}{Xiaoming Liu}, \bibinfo{person}{Zhaohan Zhang}, \bibinfo{person}{Yichen Wang}, \bibinfo{person}{Chen Liu}, \bibinfo{person}{Yu Lan}, {and} \bibinfo{person}{Chao Shen}.} \bibinfo{year}{2024}\natexlab{}.
\newblock \bibinfo{title}{Concentrate Attention: Towards Domain-Generalizable Prompt Optimization for Language Models}.
\newblock
\showeprint[arxiv]{2406.10584}~[cs.CL]
\urldef\tempurl%
\url{https://arxiv.org/abs/2406.10584}
\showURL{%
\tempurl}


\bibitem[Liang et~al\mbox{.}(2024)]%
        {liang2024cmatmultiagentcollaborationtuning}
\bibfield{author}{\bibinfo{person}{Xuechen Liang}, \bibinfo{person}{Meiling Tao}, \bibinfo{person}{Yinghui Xia}, \bibinfo{person}{Tianyu Shi}, \bibinfo{person}{Jun Wang}, {and} \bibinfo{person}{JingSong Yang}.} \bibinfo{year}{2024}\natexlab{}.
\newblock \bibinfo{title}{CMAT: A Multi-Agent Collaboration Tuning Framework for Enhancing Small Language Models}.
\newblock
\showeprint[arxiv]{2404.01663}~[cs.CL]
\urldef\tempurl%
\url{https://arxiv.org/abs/2404.01663}
\showURL{%
\tempurl}


\bibitem[Liu et~al\mbox{.}(2024)]%
        {liu-etal-2024-lost}
\bibfield{author}{\bibinfo{person}{Nelson~F. Liu}, \bibinfo{person}{Kevin Lin}, \bibinfo{person}{John Hewitt}, \bibinfo{person}{Ashwin Paranjape}, \bibinfo{person}{Michele Bevilacqua}, \bibinfo{person}{Fabio Petroni}, {and} \bibinfo{person}{Percy Liang}.} \bibinfo{year}{2024}\natexlab{}.
\newblock \showarticletitle{Lost in the Middle: How Language Models Use Long Contexts}.
\newblock \bibinfo{journal}{\emph{Transactions of the Association for Computational Linguistics}}  \bibinfo{volume}{12} (\bibinfo{year}{2024}), \bibinfo{pages}{157--173}.
\newblock
\urldef\tempurl%
\url{https://doi.org/10.1162/tacl_a_00638}
\showDOI{\tempurl}


\bibitem[MacDonald(1990)]%
        {macdonald1990one}
\bibfield{author}{\bibinfo{person}{J.~Fred MacDonald}.} \bibinfo{year}{1990}\natexlab{}.
\newblock \bibinfo{booktitle}{\emph{One Nation Under Television: The Rise and Decline of Network {TV}}}.
\newblock \bibinfo{publisher}{Pantheon Books}, \bibinfo{address}{New York}.
\newblock
\showISBNx{0394557018}


\bibitem[Markov et~al\mbox{.}(2023)]%
        {Markov2023}
\bibfield{author}{\bibinfo{person}{Todor Markov}, \bibinfo{person}{Chong Zhang}, \bibinfo{person}{Sandhini Agarwal}, \bibinfo{person}{Florentine~Eloundou Nekoul}, \bibinfo{person}{Theodore Lee}, \bibinfo{person}{Steven Adler}, \bibinfo{person}{Angela Jiang}, {and} \bibinfo{person}{Lilian Weng}.} \bibinfo{year}{2023}\natexlab{}.
\newblock \showarticletitle{A holistic approach to undesired content detection in the real world}. In \bibinfo{booktitle}{\emph{Proceedings of the Thirty-Seventh AAAI Conference on Artificial Intelligence and Thirty-Fifth Conference on Innovative Applications of Artificial Intelligence and Thirteenth Symposium on Educational Advances in Artificial Intelligence}} \emph{(\bibinfo{series}{AAAI'23/IAAI'23/EAAI'23})}. \bibinfo{publisher}{AAAI Press}, Article \bibinfo{articleno}{1683}, \bibinfo{numpages}{10}~pages.
\newblock
\showISBNx{978-1-57735-880-0}
\urldef\tempurl%
\url{https://doi.org/10.1609/aaai.v37i12.26752}
\showDOI{\tempurl}


\bibitem[Marx et~al\mbox{.}(2020)]%
        {marx20}
\bibfield{author}{\bibinfo{person}{Charles~T. Marx}, \bibinfo{person}{Flavio Du~Pin~Calmon}, {and} \bibinfo{person}{Berk Ustun}.} \bibinfo{year}{2020}\natexlab{}.
\newblock \showarticletitle{Predictive multiplicity in classification}. In \bibinfo{booktitle}{\emph{Proceedings of the 37th International Conference on Machine Learning}} \emph{(\bibinfo{series}{ICML'20})}. \bibinfo{publisher}{JMLR.org}, Article \bibinfo{articleno}{628}, \bibinfo{numpages}{10}~pages.
\newblock


\bibitem[Mikolov et~al\mbox{.}(2013)]%
        {Mikolov2013}
\bibfield{author}{\bibinfo{person}{Tomas Mikolov}, \bibinfo{person}{Kai Chen}, \bibinfo{person}{Greg~S. Corrado}, {and} \bibinfo{person}{Jeffrey Dean}.} \bibinfo{year}{2013}\natexlab{}.
\newblock \bibinfo{title}{Efficient Estimation of Word Representations in Vector Space}.
\newblock
\urldef\tempurl%
\url{http://arxiv.org/abs/1301.3781}
\showURL{%
\tempurl}


\bibitem[Mitchell et~al\mbox{.}(2021)]%
        {mitchell2021}
\bibfield{author}{\bibinfo{person}{Shira Mitchell}, \bibinfo{person}{Eric Potash}, \bibinfo{person}{Solon Barocas}, \bibinfo{person}{Alexander D’Amour}, {and} \bibinfo{person}{Kristian Lum}.} \bibinfo{year}{2021}\natexlab{}.
\newblock \showarticletitle{Algorithmic Fairness: Choices, Assumptions, and Definitions}.
\newblock \bibinfo{journal}{\emph{Annual Review of Statistics and Its Application}} \bibinfo{volume}{8}, \bibinfo{number}{1} (\bibinfo{date}{March} \bibinfo{year}{2021}), \bibinfo{pages}{141–163}.
\newblock
\showISSN{2326-831X}
\urldef\tempurl%
\url{https://doi.org/10.1146/annurev-statistics-042720-125902}
\showDOI{\tempurl}


\bibitem[{Mitra} et~al\mbox{.}(2022)]%
        {mitra2022}
\bibfield{author}{\bibinfo{person}{Debasis {Mitra}}, \bibinfo{person}{Debanjan {Mitra}}, \bibinfo{person}{Mohamed {Sabri Bensaad}}, \bibinfo{person}{Somya {Sinha}}, \bibinfo{person}{Kumud {Pant}}, \bibinfo{person}{Manu {Pant}}, \bibinfo{person}{Ankita {Priyadarshini}}, \bibinfo{person}{Pallavi {Singh}}, \bibinfo{person}{Saliha {Dassamiour}}, \bibinfo{person}{Leila {Hambaba}}, \bibinfo{person}{Periyasamy {Panneerselvam}}, {and} \bibinfo{person}{Pradeep~K. {Das Mohapatra}}.} \bibinfo{year}{2022}\natexlab{}.
\newblock \showarticletitle{{Evolution of bioinformatics and its impact on modern bio-science in the twenty-first century: Special attention to pharmacology, plant science and drug discovery}}.
\newblock \bibinfo{journal}{\emph{Computational Toxicology}}  \bibinfo{volume}{24}, Article \bibinfo{articleno}{100248} (\bibinfo{date}{Nov.} \bibinfo{year}{2022}), \bibinfo{numpages}{100248}~pages.
\newblock
\urldef\tempurl%
\url{https://doi.org/10.1016/j.comtox.2022.100248}
\showDOI{\tempurl}


\bibitem[Mullick et~al\mbox{.}(2023)]%
        {mullick-etal-2023-content}
\bibfield{author}{\bibinfo{person}{Sankha~Subhra Mullick}, \bibinfo{person}{Mohan Bhambhani}, \bibinfo{person}{Suhit Sinha}, \bibinfo{person}{Akshat Mathur}, \bibinfo{person}{Somya Gupta}, {and} \bibinfo{person}{Jidnya Shah}.} \bibinfo{year}{2023}\natexlab{}.
\newblock \showarticletitle{Content Moderation for Evolving Policies using Binary Question Answering}. In \bibinfo{booktitle}{\emph{Proceedings of the 61st Annual Meeting of the Association for Computational Linguistics (Volume 5: Industry Track)}}, \bibfield{editor}{\bibinfo{person}{Sunayana Sitaram}, \bibinfo{person}{Beata Beigman~Klebanov}, {and} \bibinfo{person}{Jason~D Williams}} (Eds.). \bibinfo{publisher}{Association for Computational Linguistics}, \bibinfo{address}{Toronto, Canada}, \bibinfo{pages}{561--573}.
\newblock
\urldef\tempurl%
\url{https://doi.org/10.18653/v1/2023.acl-industry.54}
\showDOI{\tempurl}


\bibitem[OECD(2019)]%
        {oecd2019}
\bibfield{author}{\bibinfo{person}{OECD}.} \bibinfo{year}{2019}\natexlab{}.
\newblock \bibinfo{title}{Recommendation of the Council on Artificial Intelligence}.
\newblock \bibinfo{howpublished}{OECD/LEGAL/0449}.
\newblock
\urldef\tempurl%
\url{https://legalinstruments.oecd.org/en/instruments/OECD-LEGAL-0449}
\showURL{%
\tempurl}


\bibitem[OpenAI(2023)]%
        {openai2023gpt4}
\bibfield{author}{\bibinfo{person}{OpenAI}.} \bibinfo{year}{2023}\natexlab{}.
\newblock \bibinfo{title}{{GPT-4 Technical Report}}.
\newblock \bibinfo{howpublished}{\url{https://cdn.openai.com/papers/gpt-4.pdf}}.
\newblock
\newblock
\shownote{Accessed: January 9, 2025}.


\bibitem[OpenAI(2024)]%
        {openai2024gpt4o-mini}
\bibfield{author}{\bibinfo{person}{OpenAI}.} \bibinfo{year}{2024}\natexlab{}.
\newblock \bibinfo{title}{{GPT-4o Mini: Advancing Cost-Efficient Intelligence}}.
\newblock \bibinfo{howpublished}{\url{https://openai.com/index/gpt-4o-mini-advancing-cost-efficient-intelligence/}}.
\newblock
\newblock
\shownote{Accessed: 2025-01-13}.


\bibitem[{OpenAI}(2024)]%
        {openai2024moderation}
\bibfield{author}{\bibinfo{person}{{OpenAI}}.} \bibinfo{year}{2024}\natexlab{}.
\newblock \bibinfo{title}{Using GPT-4 for Content Moderation}.
\newblock
\urldef\tempurl%
\url{https://openai.com/index/using-gpt-4-for-content-moderation/}
\showURL{%
\tempurl}
\newblock
\shownote{Accessed: 2025-01-11}.


\bibitem[Pachyderm(2021)]%
        {pachyderm2021}
\bibfield{author}{\bibinfo{person}{Inc. Pachyderm}.} \bibinfo{year}{2021}\natexlab{}.
\newblock \bibinfo{title}{Pachyderm: Version Control for Data Science and Machine Learning Pipelines}.
\newblock \bibinfo{howpublished}{\url{https://www.pachyderm.com}}.
\newblock


\bibitem[Pennington et~al\mbox{.}(2014)]%
        {pennington2014glove}
\bibfield{author}{\bibinfo{person}{Jeffrey Pennington}, \bibinfo{person}{Richard Socher}, {and} \bibinfo{person}{Christopher Manning}.} \bibinfo{year}{2014}\natexlab{}.
\newblock \showarticletitle{{G}lo{V}e: Global Vectors for Word Representation}. In \bibinfo{booktitle}{\emph{Proceedings of the 2014 Conference on Empirical Methods in Natural Language Processing ({EMNLP})}}, \bibfield{editor}{\bibinfo{person}{Alessandro Moschitti}, \bibinfo{person}{Bo~Pang}, {and} \bibinfo{person}{Walter Daelemans}} (Eds.). \bibinfo{publisher}{Association for Computational Linguistics}, \bibinfo{address}{Doha, Qatar}, \bibinfo{pages}{1532--1543}.
\newblock
\urldef\tempurl%
\url{https://doi.org/10.3115/v1/D14-1162}
\showDOI{\tempurl}


\bibitem[Perez et~al\mbox{.}(2022)]%
        {perez-etal-2022-red}
\bibfield{author}{\bibinfo{person}{Ethan Perez}, \bibinfo{person}{Saffron Huang}, \bibinfo{person}{Francis Song}, \bibinfo{person}{Trevor Cai}, \bibinfo{person}{Roman Ring}, \bibinfo{person}{John Aslanides}, \bibinfo{person}{Amelia Glaese}, \bibinfo{person}{Nat McAleese}, {and} \bibinfo{person}{Geoffrey Irving}.} \bibinfo{year}{2022}\natexlab{}.
\newblock \showarticletitle{Red Teaming Language Models with Language Models}. In \bibinfo{booktitle}{\emph{Proceedings of the 2022 Conference on Empirical Methods in Natural Language Processing}}, \bibfield{editor}{\bibinfo{person}{Yoav Goldberg}, \bibinfo{person}{Zornitsa Kozareva}, {and} \bibinfo{person}{Yue Zhang}} (Eds.). \bibinfo{publisher}{Association for Computational Linguistics}, \bibinfo{address}{Abu Dhabi, United Arab Emirates}, \bibinfo{pages}{3419--3448}.
\newblock
\urldef\tempurl%
\url{https://doi.org/10.18653/v1/2022.emnlp-main.225}
\showDOI{\tempurl}


\bibitem[Peters et~al\mbox{.}(2018)]%
        {peters-etal-2018-deep}
\bibfield{author}{\bibinfo{person}{Matthew~E. Peters}, \bibinfo{person}{Mark Neumann}, \bibinfo{person}{Mohit Iyyer}, \bibinfo{person}{Matt Gardner}, \bibinfo{person}{Christopher Clark}, \bibinfo{person}{Kenton Lee}, {and} \bibinfo{person}{Luke Zettlemoyer}.} \bibinfo{year}{2018}\natexlab{}.
\newblock \showarticletitle{Deep Contextualized Word Representations}. In \bibinfo{booktitle}{\emph{Proceedings of the 2018 Conference of the North {A}merican Chapter of the Association for Computational Linguistics: Human Language Technologies, Volume 1 (Long Papers)}}, \bibfield{editor}{\bibinfo{person}{Marilyn Walker}, \bibinfo{person}{Heng Ji}, {and} \bibinfo{person}{Amanda Stent}} (Eds.). \bibinfo{publisher}{Association for Computational Linguistics}, \bibinfo{address}{New Orleans, Louisiana}, \bibinfo{pages}{2227--2237}.
\newblock
\urldef\tempurl%
\url{https://doi.org/10.18653/v1/N18-1202}
\showDOI{\tempurl}


\bibitem[Prem and Krenn(2024)]%
        {Prem2024}
\bibfield{author}{\bibinfo{person}{Erich Prem} {and} \bibinfo{person}{Brigitte Krenn}.} \bibinfo{year}{2024}\natexlab{}.
\newblock \bibinfo{booktitle}{\emph{On Algorithmic Content Moderation}}.
\newblock \bibinfo{publisher}{Springer Nature Switzerland}, \bibinfo{address}{Cham}, \bibinfo{pages}{481--493}.
\newblock
\showISBNx{978-3-031-45304-5}
\urldef\tempurl%
\url{https://doi.org/10.1007/978-3-031-45304-5_30}
\showDOI{\tempurl}


\bibitem[Reda(2017)]%
        {Reda2017}
\bibfield{author}{\bibinfo{person}{Felix Reda}.} \bibinfo{year}{2017}\natexlab{}.
\newblock \bibinfo{title}{When filters fail: These cases show we can’t trust algorithms to clean up the internet}.
\newblock
\urldef\tempurl%
\url{https://felixreda.eu/2017/09/when-filters-fail/}
\showURL{%
\tempurl}
\newblock
\shownote{Accessed: 2025-01-09}.


\bibitem[{Reddit, Inc.}(2025)]%
        {reddit_content_policy}
\bibfield{author}{\bibinfo{person}{{Reddit, Inc.}}} \bibinfo{year}{2025}\natexlab{}.
\newblock \bibinfo{title}{Reddit Content Policy}.
\newblock
\urldef\tempurl%
\url{https://redditinc.com/policies/content-policy}
\showURL{%
\tempurl}
\newblock
\shownote{Accessed: 2025-01-11}.


\bibitem[Roberts(2019)]%
        {roberts2019}
\bibfield{author}{\bibinfo{person}{Sarah~T. Roberts}.} \bibinfo{year}{2019}\natexlab{}.
\newblock \bibinfo{booktitle}{\emph{Behind the Screen: Content Moderation in the Shadows of Social Media}}.
\newblock \bibinfo{publisher}{Yale University Press}.
\newblock
\showISBNx{roberts2019}
\urldef\tempurl%
\url{http://www.jstor.org/stable/j.ctvhrcz0v}
\showURL{%
\tempurl}


\bibitem[Sahoo et~al\mbox{.}(2024)]%
        {sahoo2024systematicsurveypromptengineering}
\bibfield{author}{\bibinfo{person}{Pranab Sahoo}, \bibinfo{person}{Ayush~Kumar Singh}, \bibinfo{person}{Sriparna Saha}, \bibinfo{person}{Vinija Jain}, \bibinfo{person}{Samrat Mondal}, {and} \bibinfo{person}{Aman Chadha}.} \bibinfo{year}{2024}\natexlab{}.
\newblock \bibinfo{title}{A Systematic Survey of Prompt Engineering in Large Language Models: Techniques and Applications}.
\newblock
\showeprint[arxiv]{2402.07927}~[cs.AI]
\urldef\tempurl%
\url{https://arxiv.org/abs/2402.07927}
\showURL{%
\tempurl}


\bibitem[Sandri et~al\mbox{.}(2023)]%
        {sandri-etal-2023-dont}
\bibfield{author}{\bibinfo{person}{Marta Sandri}, \bibinfo{person}{Elisa Leonardelli}, \bibinfo{person}{Sara Tonelli}, {and} \bibinfo{person}{Elisabetta Jezek}.} \bibinfo{year}{2023}\natexlab{}.
\newblock \showarticletitle{Why Don`t You Do It Right? Analysing Annotators' Disagreement in Subjective Tasks}. In \bibinfo{booktitle}{\emph{Proceedings of the 17th Conference of the European Chapter of the Association for Computational Linguistics}}, \bibfield{editor}{\bibinfo{person}{Andreas Vlachos} {and} \bibinfo{person}{Isabelle Augenstein}} (Eds.). \bibinfo{publisher}{Association for Computational Linguistics}, \bibinfo{address}{Dubrovnik, Croatia}, \bibinfo{pages}{2428--2441}.
\newblock
\urldef\tempurl%
\url{https://doi.org/10.18653/v1/2023.eacl-main.178}
\showDOI{\tempurl}


\bibitem[Schulhoff et~al\mbox{.}(2024)]%
        {schulhoff2024}
\bibfield{author}{\bibinfo{person}{Sander Schulhoff}, \bibinfo{person}{Michael Ilie}, \bibinfo{person}{Nishant Balepur}, \bibinfo{person}{Konstantine Kahadze}, \bibinfo{person}{Amanda Liu}, \bibinfo{person}{Chenglei Si}, \bibinfo{person}{Yinheng Li}, \bibinfo{person}{Aayush Gupta}, \bibinfo{person}{HyoJung Han}, \bibinfo{person}{Sevien Schulhoff}, \bibinfo{person}{Pranav~Sandeep Dulepet}, \bibinfo{person}{Saurav Vidyadhara}, \bibinfo{person}{Dayeon Ki}, \bibinfo{person}{Sweta Agrawal}, \bibinfo{person}{Chau Pham}, \bibinfo{person}{Gerson~C. Kroiz}, \bibinfo{person}{Feileen Li}, \bibinfo{person}{Hudson Tao}, \bibinfo{person}{Ashay Srivastava}, \bibinfo{person}{Hevander~Da Costa}, \bibinfo{person}{Saloni Gupta}, \bibinfo{person}{Megan~L. Rogers}, \bibinfo{person}{Inna Goncearenco}, \bibinfo{person}{Giuseppe Sarli}, \bibinfo{person}{Igor Galynker}, \bibinfo{person}{Denis Peskoff}, \bibinfo{person}{Marine Carpuat}, \bibinfo{person}{Jules White}, \bibinfo{person}{Shyamal Anadkat},
  \bibinfo{person}{Alexander~Miserlis Hoyle}, {and} \bibinfo{person}{Philip Resnik}.} \bibinfo{year}{2024}\natexlab{}.
\newblock \showarticletitle{The Prompt Report: A Systematic Survey of Prompting Techniques}.
\newblock \bibinfo{journal}{\emph{CoRR}}  \bibinfo{volume}{abs/2406.06608} (\bibinfo{year}{2024}).
\newblock
\urldef\tempurl%
\url{https://doi.org/10.48550/arXiv.2406.06608}
\showURL{%
\tempurl}


\bibitem[Sclar et~al\mbox{.}(2024)]%
        {sclar2024quantifying}
\bibfield{author}{\bibinfo{person}{Melanie Sclar}, \bibinfo{person}{Yejin Choi}, \bibinfo{person}{Yulia Tsvetkov}, {and} \bibinfo{person}{Alane Suhr}.} \bibinfo{year}{2024}\natexlab{}.
\newblock \showarticletitle{Quantifying Language Models' Sensitivity to Spurious Features in Prompt Design or: How I learned to start worrying about prompt formatting}. In \bibinfo{booktitle}{\emph{The Twelfth International Conference on Learning Representations}}.
\newblock
\urldef\tempurl%
\url{https://openreview.net/forum?id=RIu5lyNXjT}
\showURL{%
\tempurl}


\bibitem[Simson et~al\mbox{.}(2024)]%
        {simson2024}
\bibfield{author}{\bibinfo{person}{Jan Simson}, \bibinfo{person}{Florian Pfisterer}, {and} \bibinfo{person}{Christoph Kern}.} \bibinfo{year}{2024}\natexlab{}.
\newblock \showarticletitle{One Model Many Scores: Using Multiverse Analysis to Prevent Fairness Hacking and Evaluate the Influence of Model Design Decisions}. In \bibinfo{booktitle}{\emph{Proceedings of the 2024 ACM Conference on Fairness, Accountability, and Transparency}} (Rio de Janeiro, Brazil) \emph{(\bibinfo{series}{FAccT '24})}. \bibinfo{publisher}{Association for Computing Machinery}, \bibinfo{address}{New York, NY, USA}, \bibinfo{pages}{1305–1320}.
\newblock
\showISBNx{9798400704505}
\urldef\tempurl%
\url{https://doi.org/10.1145/3630106.3658974}
\showDOI{\tempurl}


\bibitem[Singh et~al\mbox{.}(2024)]%
        {singh2024rethinking}
\bibfield{author}{\bibinfo{person}{Chandan Singh}, \bibinfo{person}{Jeevana~Priya Inala}, \bibinfo{person}{Michel Galley}, \bibinfo{person}{Rich Caruana}, {and} \bibinfo{person}{Jianfeng Gao}.} \bibinfo{year}{2024}\natexlab{}.
\newblock \showarticletitle{Rethinking Interpretability in the Era of Large Language Models}.
\newblock \bibinfo{journal}{\emph{arXiv preprint arXiv:2402.01761}} (\bibinfo{year}{2024}).
\newblock


\bibitem[Singhal et~al\mbox{.}(2023)]%
        {Singhal_2023}
\bibfield{author}{\bibinfo{person}{Mohit Singhal}, \bibinfo{person}{Chen Ling}, \bibinfo{person}{Nihal Kumarswamy}, \bibinfo{person}{Gianluca Stringhini}, {and} \bibinfo{person}{Shirin Nilizadeh}.} \bibinfo{year}{2023}\natexlab{}.
\newblock \showarticletitle{SoK: Content Moderation in Social Media, from Guidelines to Enforcement, and Research to Practice}.
\newblock \bibinfo{journal}{\emph{2023 IEEE 8th European Symposium on Security and Privacy (EuroS\&P)}} (\bibinfo{year}{2023}), \bibinfo{pages}{868--895}.
\newblock


\bibitem[Spinellis(2005)]%
        {spinellis2005version}
\bibfield{author}{\bibinfo{person}{Diomidis Spinellis}.} \bibinfo{year}{2005}\natexlab{}.
\newblock \showarticletitle{Version control systems}.
\newblock \bibinfo{journal}{\emph{IEEE Software}} \bibinfo{volume}{22}, \bibinfo{number}{5} (\bibinfo{year}{2005}), \bibinfo{pages}{108--109}.
\newblock
\urldef\tempurl%
\url{https://doi.org/10.1109/MS.2005.145}
\showDOI{\tempurl}


\bibitem[Suzgun and Kalai(2024)]%
        {suzgun2024metapromptingenhancinglanguagemodels}
\bibfield{author}{\bibinfo{person}{Mirac Suzgun} {and} \bibinfo{person}{Adam~Tauman Kalai}.} \bibinfo{year}{2024}\natexlab{}.
\newblock \bibinfo{title}{Meta-Prompting: Enhancing Language Models with Task-Agnostic Scaffolding}.
\newblock
\showeprint[arxiv]{2401.12954}~[cs.CL]
\urldef\tempurl%
\url{https://arxiv.org/abs/2401.12954}
\showURL{%
\tempurl}


\bibitem[Touvron et~al\mbox{.}(2023)]%
        {touvron2023llama2}
\bibfield{author}{\bibinfo{person}{Hugo Touvron}, \bibinfo{person}{Louis Martin}, \bibinfo{person}{Kevin Stone}, \bibinfo{person}{Patrick Albert}, \bibinfo{person}{Amjad Almahairi}, \bibinfo{person}{Yasaman Babaei}, \bibinfo{person}{...}, {and} \bibinfo{person}{Daniel Jurafsky}.} \bibinfo{year}{2023}\natexlab{}.
\newblock \showarticletitle{Llama 2: Open Foundation and Fine-Tuned Chat Models}.
\newblock \bibinfo{journal}{\emph{arXiv preprint arXiv:2307.09288}} (\bibinfo{year}{2023}).
\newblock


\bibitem[van Dijck(2013)]%
        {vandijck2013culture}
\bibfield{author}{\bibinfo{person}{Jos{\'e} van Dijck}.} \bibinfo{year}{2013}\natexlab{}.
\newblock \bibinfo{booktitle}{\emph{The Culture of Connectivity: A Critical History of Social Media}}.
\newblock \bibinfo{publisher}{Oxford University Press}, \bibinfo{address}{Oxford}.
\newblock


\bibitem[Vaswani et~al\mbox{.}(2017)]%
        {Vaswani2017}
\bibfield{author}{\bibinfo{person}{Ashish Vaswani}, \bibinfo{person}{Noam Shazeer}, \bibinfo{person}{Niki Parmar}, \bibinfo{person}{Jakob Uszkoreit}, \bibinfo{person}{Llion Jones}, \bibinfo{person}{Aidan~N Gomez}, \bibinfo{person}{\L~ukasz Kaiser}, {and} \bibinfo{person}{Illia Polosukhin}.} \bibinfo{year}{2017}\natexlab{}.
\newblock \showarticletitle{Attention is All you Need}. In \bibinfo{booktitle}{\emph{Advances in Neural Information Processing Systems}}, \bibfield{editor}{\bibinfo{person}{I.~Guyon}, \bibinfo{person}{U.~Von Luxburg}, \bibinfo{person}{S.~Bengio}, \bibinfo{person}{H.~Wallach}, \bibinfo{person}{R.~Fergus}, \bibinfo{person}{S.~Vishwanathan}, {and} \bibinfo{person}{R.~Garnett}} (Eds.), Vol.~\bibinfo{volume}{30}. \bibinfo{publisher}{Curran Associates, Inc.}
\newblock
\urldef\tempurl%
\url{https://proceedings.neurips.cc/paper_files/paper/2017/file/3f5ee243547dee91fbd053c1c4a845aa-Paper.pdf}
\showURL{%
\tempurl}


\bibitem[Verma and Rubin(2018)]%
        {verma2018}
\bibfield{author}{\bibinfo{person}{Sahil Verma} {and} \bibinfo{person}{Julia Rubin}.} \bibinfo{year}{2018}\natexlab{}.
\newblock \showarticletitle{Fairness definitions explained}. In \bibinfo{booktitle}{\emph{2018 IEEE/ACM International Workshop on Software Fairness (FairWare)}}. IEEE, \bibinfo{pages}{1--7}.
\newblock


\bibitem[Wang et~al\mbox{.}(2024)]%
        {wang-etal-2024-unleashing}
\bibfield{author}{\bibinfo{person}{Zhenhailong Wang}, \bibinfo{person}{Shaoguang Mao}, \bibinfo{person}{Wenshan Wu}, \bibinfo{person}{Tao Ge}, \bibinfo{person}{Furu Wei}, {and} \bibinfo{person}{Heng Ji}.} \bibinfo{year}{2024}\natexlab{}.
\newblock \showarticletitle{Unleashing the Emergent Cognitive Synergy in Large Language Models: A Task-Solving Agent through Multi-Persona Self-Collaboration}. In \bibinfo{booktitle}{\emph{Proceedings of the 2024 Conference of the North American Chapter of the Association for Computational Linguistics: Human Language Technologies (Volume 1: Long Papers)}}, \bibfield{editor}{\bibinfo{person}{Kevin Duh}, \bibinfo{person}{Helena Gomez}, {and} \bibinfo{person}{Steven Bethard}} (Eds.). \bibinfo{publisher}{Association for Computational Linguistics}, \bibinfo{address}{Mexico City, Mexico}, \bibinfo{pages}{257--279}.
\newblock
\urldef\tempurl%
\url{https://doi.org/10.18653/v1/2024.naacl-long.15}
\showDOI{\tempurl}


\bibitem[Wei et~al\mbox{.}(2022)]%
        {wei2022}
\bibfield{author}{\bibinfo{person}{Jason Wei}, \bibinfo{person}{Xuezhi Wang}, \bibinfo{person}{Dale Schuurmans}, \bibinfo{person}{Maarten Bosma}, \bibinfo{person}{Brian Ichter}, \bibinfo{person}{Fei Xia}, \bibinfo{person}{Ed~H. Chi}, \bibinfo{person}{Quoc~V. Le}, {and} \bibinfo{person}{Denny Zhou}.} \bibinfo{year}{2022}\natexlab{}.
\newblock \showarticletitle{Chain-of-thought prompting elicits reasoning in large language models}. In \bibinfo{booktitle}{\emph{Proceedings of the 36th International Conference on Neural Information Processing Systems}} (New Orleans, LA, USA) \emph{(\bibinfo{series}{NIPS '22})}. \bibinfo{address}{Red Hook, NY, USA}, Article \bibinfo{articleno}{1800}, \bibinfo{numpages}{14}~pages.
\newblock


\bibitem[Winner(1980)]%
        {winner1980artifacts}
\bibfield{author}{\bibinfo{person}{Langdon Winner}.} \bibinfo{year}{1980}\natexlab{}.
\newblock \showarticletitle{Do Artifacts Have Politics?}
\newblock \bibinfo{journal}{\emph{Daedalus}} \bibinfo{volume}{109}, \bibinfo{number}{1} (\bibinfo{year}{1980}), \bibinfo{pages}{121--136}.
\newblock


\bibitem[y~Arcas(2022)]%
        {blaise2022}
\bibfield{author}{\bibinfo{person}{Blaise~Agüera y Arcas}.} \bibinfo{year}{2022}\natexlab{}.
\newblock \showarticletitle{Do Large Language Models Understand Us?}
\newblock \bibinfo{journal}{\emph{Daedalus}} \bibinfo{volume}{151}, \bibinfo{number}{2} (\bibinfo{date}{05} \bibinfo{year}{2022}), \bibinfo{pages}{183--197}.
\newblock
\showISSN{0011-5266}
\urldef\tempurl%
\url{https://doi.org/10.1162/daed_a_01909}
\showDOI{\tempurl}


\bibitem[Yao et~al\mbox{.}(2023)]%
        {yao2023react}
\bibfield{author}{\bibinfo{person}{Shunyu Yao}, \bibinfo{person}{Jeffrey Zhao}, \bibinfo{person}{Dian Yu}, \bibinfo{person}{Nan Du}, \bibinfo{person}{Izhak Shafran}, {and} \bibinfo{person}{Karthik Narasimhan}.} \bibinfo{year}{2023}\natexlab{}.
\newblock \showarticletitle{ReAct: Synergizing reasoning and acting in language models}. In \bibinfo{booktitle}{\emph{International Conference on Learning Representations}}.
\newblock


\bibitem[Yeung(2019)]%
        {yeung2019algorithmic}
\bibfield{author}{\bibinfo{person}{Karen Yeung}.} \bibinfo{year}{2019}\natexlab{}.
\newblock \showarticletitle{Algorithmic Regulation: A Critical Interrogation}.
\newblock \bibinfo{journal}{\emph{Regulation \& Governance}} \bibinfo{volume}{13}, \bibinfo{number}{1} (\bibinfo{year}{2019}), \bibinfo{pages}{94--114}.
\newblock
\urldef\tempurl%
\url{https://doi.org/10.1111/rego.12158}
\showDOI{\tempurl}


\bibitem[Zhang et~al\mbox{.}(2024)]%
        {zhang2024attentioninstructionamplifyingattention}
\bibfield{author}{\bibinfo{person}{Meiru Zhang}, \bibinfo{person}{Zaiqiao Meng}, {and} \bibinfo{person}{Nigel Collier}.} \bibinfo{year}{2024}\natexlab{}.
\newblock \bibinfo{title}{Attention Instruction: Amplifying Attention in the Middle via Prompting}.
\newblock
\showeprint[arxiv]{2406.17095}~[cs.CL]
\urldef\tempurl%
\url{https://arxiv.org/abs/2406.17095}
\showURL{%
\tempurl}


\end{thebibliography}

\appendix

\begin{figure}[ht!]
    \centering
    \includegraphics[width=0.8\textwidth]{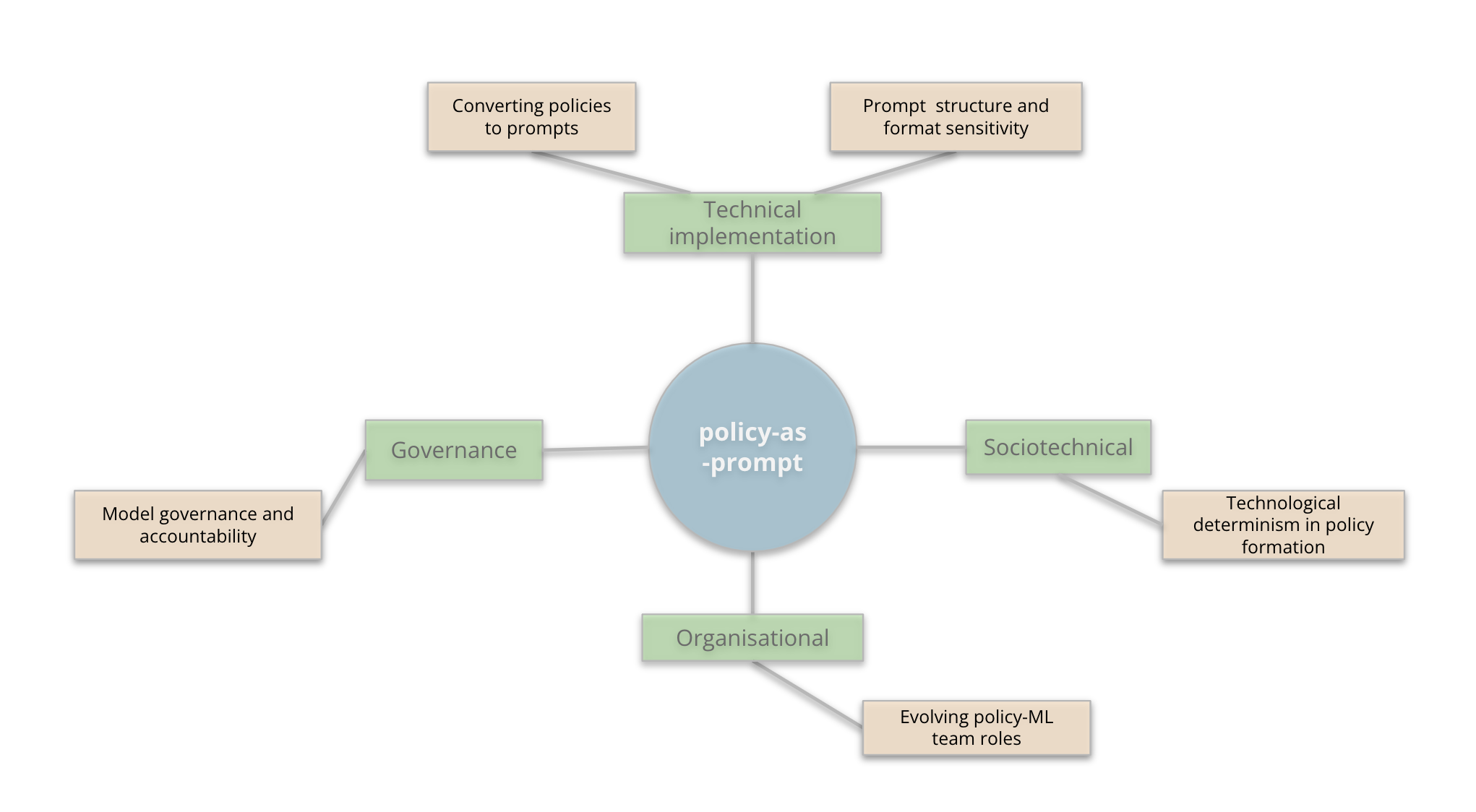}
    \caption{\textbf{Challenges} (light orange) across different \textbf{areas} (green) in ‘policy-as-prompt’ implementation.}
    \Description{Diagram illustrating challenges in implementing the concept of 'policy-as-prompt,' organized into four domains: Technical Implementation, Sociotechnical, Organisational, and Governance. Each domain connects to specific challenges, such as 'Converting policies to prompts' and 'Prompt structure and format sensitivity' under Technical Implementation, 'Technical determinism in policy formation' under Sociotechnical, 'Evolving policy-ML team roles' under Organisational, and 'Model governance and accountability' under Governance. The central node, 'policy-as-prompt,' anchors the relationships between these areas.}
    \label{fig:spider}
\end{figure}

\section{Experiments}
\label{sec:exp}
\subsection{Prompt style variations}

\begin{figure}[!ht]

\centering

\begin{minipage}[t]{0.18\textwidth} 
\textbf{1. Punctuated}
\end{minipage}
\begin{minipage}[t]{0.73\textwidth} 
\begin{tcolorbox}[colback=white, arc=10pt, boxrule=0.5pt]
\footnotesize 
\textnormal{Pr\ \ \ mpts that incite violence or hatred towards a, person or group, of people based on race religion gender identity or expression sex, ethnicity,, nationality sexual orientation veteran, status age disability or other characteristics associa\ \ \ ed with systemic discrimination or marginalization includes but may not be limited to: prai\ \ \ sing supporting, or calling for violence or exclusion, against a person or group of people based on the characteristics lis \ ted above, dehumanizing statements... }
\end{tcolorbox}
\end{minipage}

\vspace{0.3cm} 

\begin{minipage}[t]{0.18\textwidth}
\textbf{2. Structured} 
\end{minipage}
\begin{minipage}[t]{0.73\textwidth}
\begin{tcolorbox}[colback=white, arc=10pt, boxrule=0.5pt]
\footnotesize  
\textnormal{
- Prompts that incite violence or hatred towards individuals or groups are prohibited. 
- These prompts target characteristics such as: 
 \begin{itemize}
  \item[-] race 
  \item[-] religion  
  \item[-] gender identity or expression  
  \item[-] sex  
  \item[-] ethnicity  
  \item[-] nationality  
  \item[-] sexual orientation  
  \item[-] veteran status  
  \item[-] age  
  \item[-] disability  
  \item[-] other characteristics linked with systemic discrimination or marginalization  
 \end{itemize}
- Prohibited actions include, but are not limited to:  
 \begin{itemize}
     \item[-] Praising, supporting, or calling for violence or exclusion against individuals or groups based on the aforementioned characteristics.   
     \item[-] Making dehumanizing statements...
 \end{itemize}
}
\end{tcolorbox}
\end{minipage}

\vspace{0.3cm} 

\begin{minipage}[t]{0.18\textwidth}
\textbf{3. Concise}
\end{minipage}
\begin{minipage}[t]{0.73\textwidth}
\begin{tcolorbox}[colback=white, arc=10pt, boxrule=0.5pt]
\footnotesize 
\textnormal{Policy that prohibits content inciting violence or hatred on grounds of race, religion, gender identity, sex, ethnicity, nationality, sexual orientation, veteran status, age, disability, or other marginalized traits includes, but is not limited to: endorsing or advocating violence or exclusion, dehumanizing statements...}
\end{tcolorbox}
\end{minipage}

\vspace{0.3cm} 

\begin{minipage}[t]{0.18\textwidth}
\textbf{4. Verbose} 
\end{minipage}
\begin{minipage}[t]{0.73\textwidth}
\begin{tcolorbox}[colback=white, arc=10pt, boxrule=0.5pt]
\footnotesize 
\textnormal{The policy is dedicated to addressing and regulating prompts that have the potential to incite or provoke acts of violence or expressions of hatred directed toward an individual or a group of individuals. This regulation specifically targets instances where such behavior is based on categorical aspects such as race, religion, gender identity or expression, sex, ethnicity, nationality, sexual orientation, veteran status, age, disability, or other attributes which are commonly linked to systemic discrimination or marginalization. The policy encompasses, but is not strictly limited to, behaviors such as: endorsing, supporting, or advocating for acts of violence or exclusion toward an individual or groups of individuals based on any of the aforementioned characteristics, making statements that dehumanize an...}
\end{tcolorbox}
\end{minipage}

\vspace{0.3cm} 

\begin{minipage}[t]{0.18\textwidth}
\textbf{5. Annotator} 
\end{minipage}
\begin{minipage}[t]{0.73\textwidth}
\begin{tcolorbox}[colback=white, arc=10pt, boxrule=0.5pt]
\footnotesize 
\textnormal{If a message encourages violence or hatred towards someone because of their race, religion, gender identity, sex, ethnicity, nationality, sexual orientation, veteran status, age, disability, or other similar traits, it's not allowed. This includes actions like praising, supporting, or demanding violence or exclusion of these individuals. It also involves saying dehumanizing things...} 
\end{tcolorbox}
\end{minipage}

\caption{The illustration of different prompt styles for hate speech policy.}
 \Description{}
\label{fig:prompt-templates}
\end{figure}


\begin{figure}[h]
  \centering
  \begin{subfigure}[t]{0.48\linewidth}
    \centering
    \includegraphics[width=\linewidth]{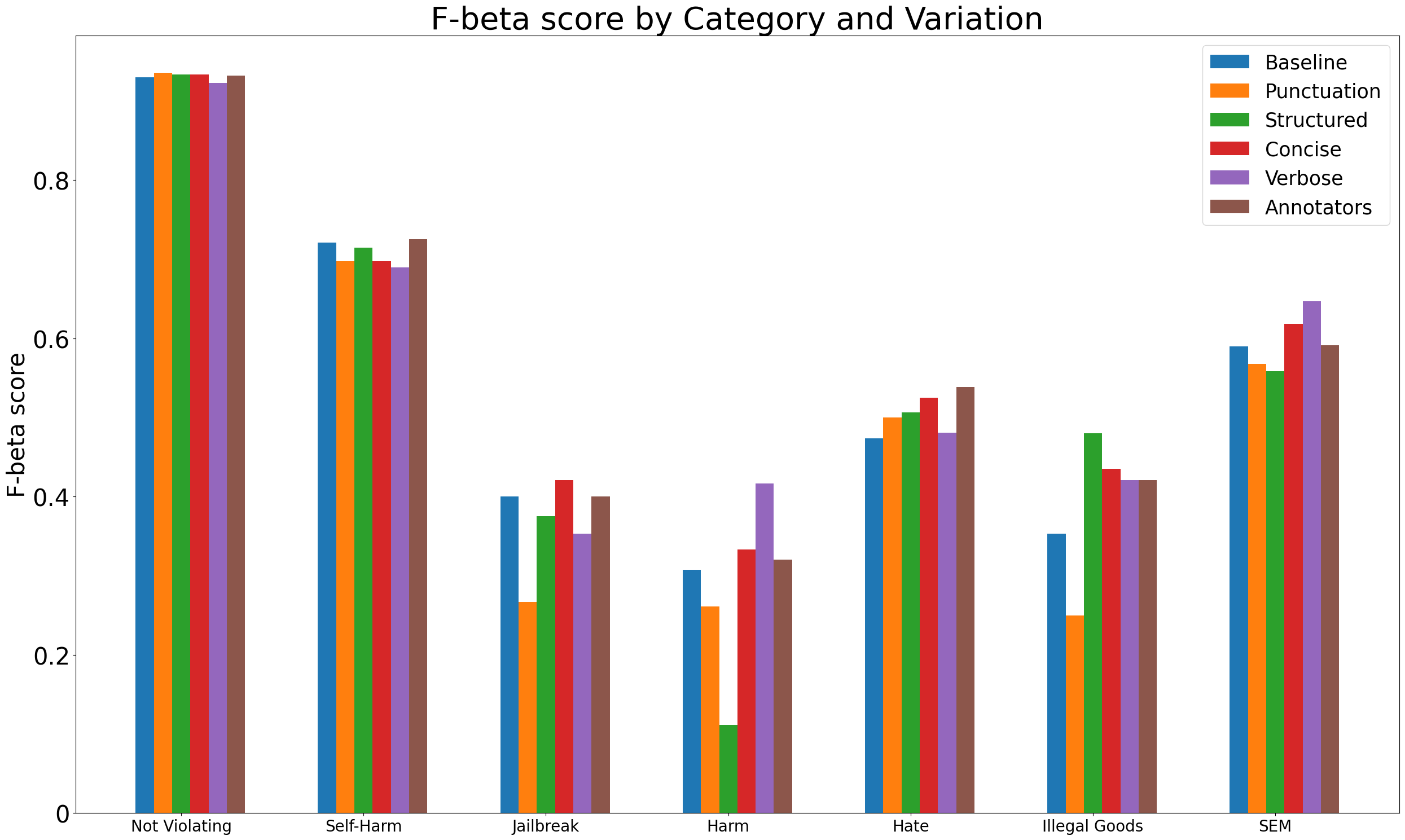}
    \caption{F-beta scores for the different textual variations considered, across some of the most important categories.}
    \Description{}
    \label{fig:performance_exp1}
  \end{subfigure}
  \hfill
  \begin{subfigure}[t]{0.48\linewidth}
    \centering
    \includegraphics[width=\linewidth]{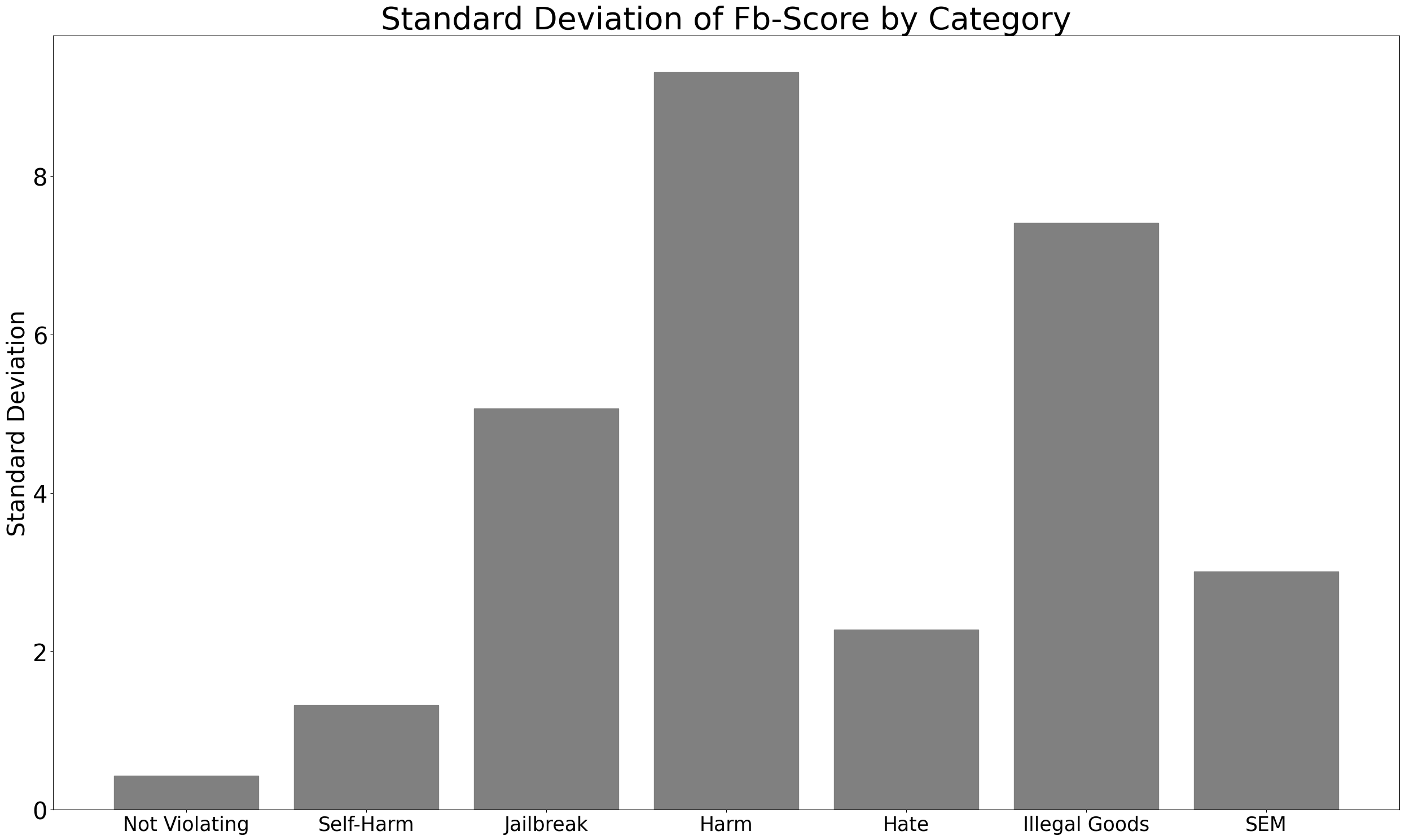}
    \caption{Standard Deviation for the performance of the main policies considered.}
    \Description{}
    \label{fig:sd_per_category}
  \end{subfigure}
  \caption{Comparison of performance metrics across categories and their standard deviations.}
  \label{fig:combined_std}
\end{figure}


We observe that model performance, as measured by the F-beta score, fluctuates across textual prompt variations (see Figure~\ref{fig:prompt-templates} for an example of the template variations and Figure~\ref{fig:performance_exp1} for the results). Categories such as Not Violating and Self-Harm maintain relatively high and stable scores, while more contentious categories like Jailbreak, Harm, and Illegal Goods display marked sensitivity. This disparity highlights that model performance is prone to degradation when subjected to structural changes in the policy prompt.  This directly supports the thesis of Section 3.2, highlighting the role of prompt design in shaping moderation outcomes. The variations in F-beta scores (Figure \ref{fig:performance_exp1}) and the high standard deviation (Figure \ref{fig:sd_per_category}) demonstrate predictive multiplicity in action. Even though the overall performance across prompt variations may appear statistically comparable, the conflicting behaviours observed in key categories (e.g., Jailbreak, Harm) point to a deeper underlying issue: different prompt structures lead to divergent model predictions for the same input samples.

\subsection{Policy Evolution Snapshots} 
To examine the role of policy specificity, we created temporal snapshots of policy descriptions, starting with minimal drafts (one-line descriptions) and progressively adding detail to arrive at comprehensive guidelines (as seen in Table \ref{tab:table_temporal} reversely).

\setlength{\arrayrulewidth}{1pt} 

\begin{table}[h!]
\centering
\setlength{\arrayrulewidth}{1pt} 
\begin{tabularx}{\textwidth}{>{\raggedright\arraybackslash}p{0.2\textwidth} 
                                   >{\raggedright\arraybackslash}p{0.18\textwidth} 
                                   >{\raggedright\arraybackslash}p{0.18\textwidth} 
                                   >{\raggedright\arraybackslash}p{0.18\textwidth} 
                                   >{\raggedright\arraybackslash}p{0.18\textwidth}}
\textnormal{\fontsize{6}{8}\selectfont Prompts that incite violence or hatred towards a person or group of people based on race, religion, gender identity or expression, sex, ethnicity, nationality, sexual orientation, veteran status, age, disability or other characteristics associated with systemic discrimination or marginalization includes, but may not be limited to: - praising, supporting, or calling for violence or exclusion against a person or group of people based on the characteristics listed above - dehumanizing statements about a person or group based on the protected characteristics listed above - using hateful slurs when targeting someone/a group of people based on their protected characteristic promoting or glorifying hate groups/hate figures and their associated images, and/or symbols.} & 
\textnormal{\fontsize{6}{8}\selectfont Prompts that incite violence or hatred towards a person or group of people based on race, religion, gender identity or expression, sex, ethnicity, nationality, sexual orientation, veteran status, age, disability or other characteristics associated with systemic discrimination or marginalization includes, but may not be limited to: - praising, supporting, or calling for violence or exclusion against a person or group of people based on the characteristics listed above - dehumanizing statements about a person or group based on the protected characteristics listed above.} & 
\textnormal{\fontsize{6}{8}\selectfont Prompts that incite violence or hatred towards a person or group of people based on race, religion, gender identity or expression, sex, ethnicity, nationality, sexual orientation, veteran status, age, disability or other characteristics associated with systemic discrimination or marginalization includes, but may not be limited to: - praising, supporting, or calling for violence against a person or group of people based on the characteristics listed above.} & 
\textnormal{\fontsize{6}{8}\selectfont Prompts that incite violence or hatred towards a person or group of people based on race, religion, gender identity or expression, sex, ethnicity, nationality, sexual orientation, veteran status, age, disability or other characteristics associated with systemic discrimination or marginalization includes using hateful slurs when targeting someone/a group of people based on their protected characteristic.} & 
\textnormal{\fontsize{6}{8}\selectfont Prompts that incite violence or hatred towards a person or group of people based on race, religion, gender identity or expression, sex, ethnicity, nationality, sexual orientation, veteran status, age, disability or other characteristics associated with systemic discrimination or marginalization.} \\[10pt]
\multicolumn{1}{c}{\fontsize{8}{10}\selectfont Snapshot 0 } & 
\multicolumn{1}{c}{\fontsize{8}{10}\selectfont Snapshot 1 } & 
\multicolumn{1}{c}{\fontsize{8}{10}\selectfont Snapshot 2 } & 
\multicolumn{1}{c}{\fontsize{8}{10}\selectfont Snapshot 3} & 
\multicolumn{1}{c}{\fontsize{8}{10}\selectfont Snapshot 4} \\
\end{tabularx}
\caption{\fontsize{8}{10}\selectfont Different temporal snapshots for the evolution of policy category: Hate.}
\label{tab:table_temporal}
\end{table}

\subsection{Model Size}

\begin{figure}[h]
  \centering
  \begin{subfigure}[t]{0.48\linewidth}
    \centering
    \includegraphics[width=\linewidth]{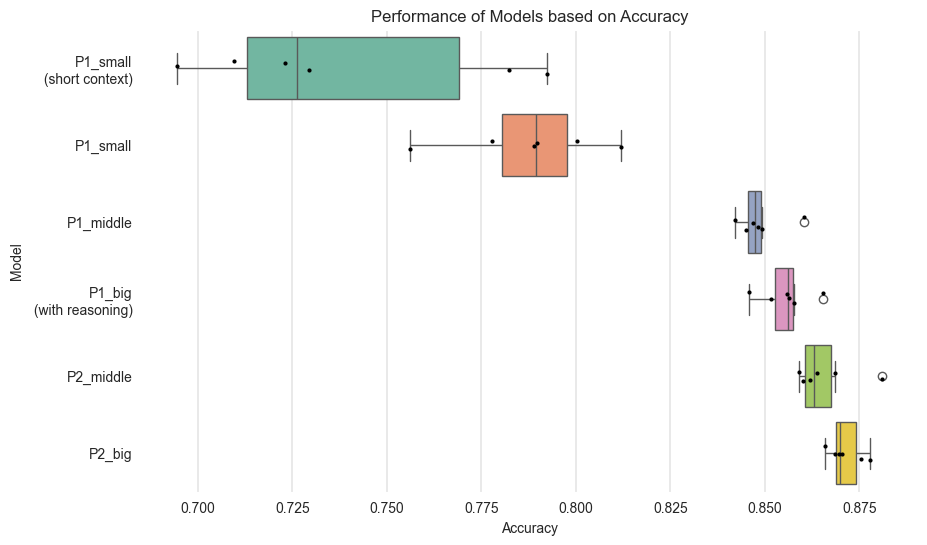}
    \caption{P}
    \Description{}
    \label{fig:model_size_variation}
  \end{subfigure}
  \hfill
  \begin{subfigure}[t]{0.48\linewidth}
    \centering
    \includegraphics[width=\linewidth]{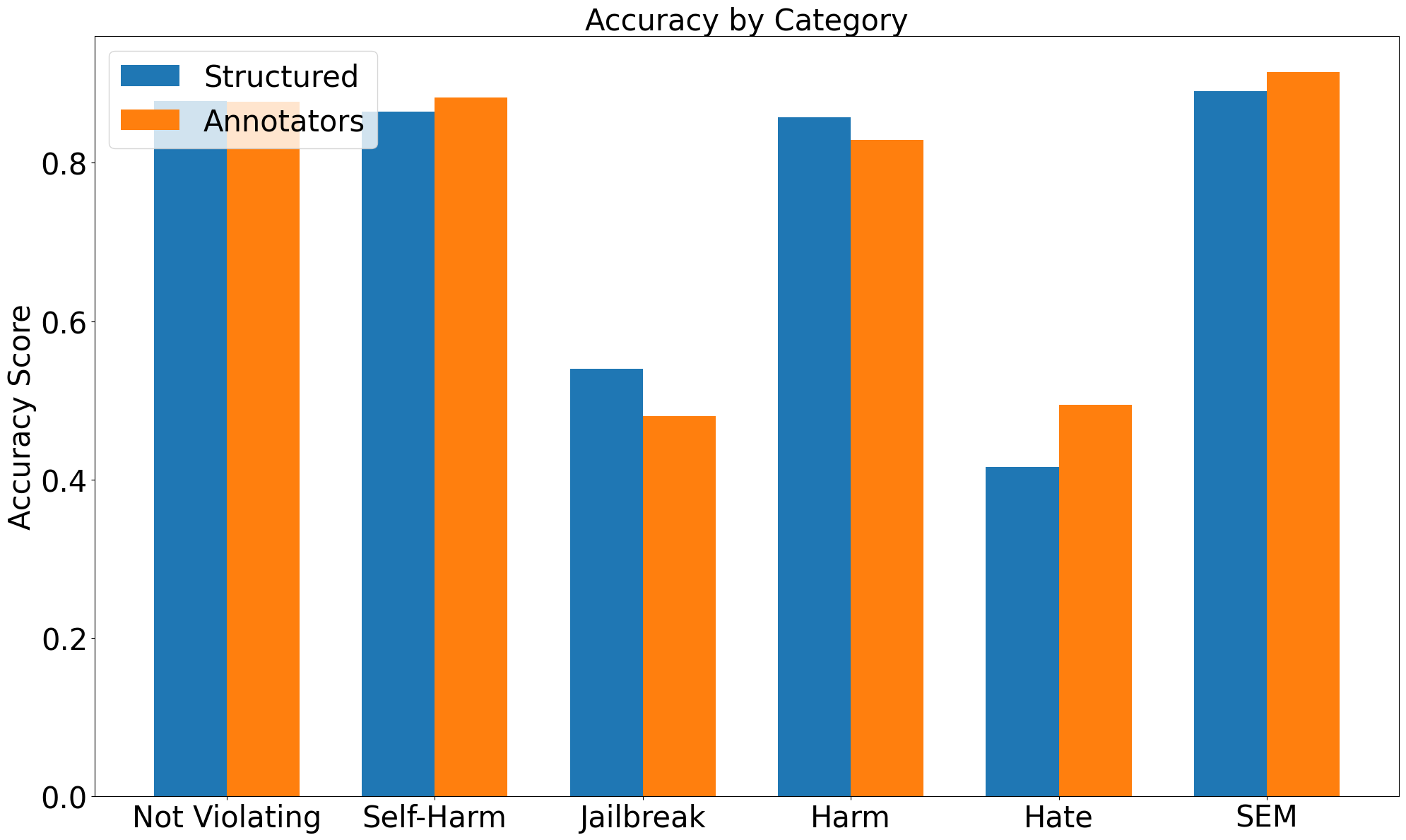}
    \caption{}
    \Description{}
    \label{fig:topmodel_per_category}
  \end{subfigure}
  \caption{Analysing the effect of model provider and size in sensitivity and predictive multiplicity. a) The performance distribution of different model sizes for the six prompt template variations. b) Accuracy across policy categories for the `Structured' and `Annotators' prompt types.}
  \label{fig:combined}
\end{figure}

To gain insight how models with different size respond to variations of the prompt, we ran experiments using models with different size from different providers. In Figure \ref{fig:model_size_variation} we plot the performance distribution and use `P\#' to indicate different model provider series. Middle sized models are between the 32B and the 64B parameters mark, while big ones surpass the 100B. Overall, while we observe that the sensitivity to prompt structure and style decreases with model size, it is still present but to a lesser degree. We also observe that different providers - thus different pre-training strategies- affect the model sensitivity. Still, even for the most capable models, predictive multiplicity arises  as can be seen in \ref{fig:topmodel_per_category}, where the variants \textit{Structured} and \textit{Annotator-Friendly} present significant performance divergences for certain categories, despite negligible differences in averaged accuracy. For example, accuracy scores for Hate are around five points higher for the \textit{Annotator-Friendly} variation. 

\end{document}